\documentclass[aps,onecolumn,nofootinbib,floatfix,letterpaper,showpacs,notitlepage]{revtex4}

\usepackage{times}
\usepackage{graphicx}
\usepackage{epsfig}
\usepackage{subfigure}
\usepackage{amssymb,amsmath}
\usepackage{hyperref}
\usepackage{setspace}
\usepackage{url}

\def\tst{\tilde t}
\def\ttau{\tilde \tau}

\hypersetup{
    colorlinks = true,
    linkcolor = blue,
    anchorcolor = blue,
    citecolor = blue,
    filecolor = blue,
    pagecolor = blue,
    urlcolor = blue
}

\begin{document}

\title{
Particle Spectroscopy of Supersymmetric SO(10) 
with Non-Universal Gaugino Masses 
}
\author{Nobuchika Okada~$^{a}$}
\email{okadan@ua.edu.}
\author{Shabbar Raza~$^{b}$}
\email{shabbar@udel.edu. On study leave from: Department of Physics, FUUAST, Islamabad, Pakistan.}
\author{Qaisar Shafi~$^{b}$}
\affiliation{
$^{a}$~ Department of Physics and Astronomy, 
University of Alabama, Tuscaloosa, AL 35487, USA \\
$^{b}$Bartol Research Institute, 
Department of Physics and Astronomy, 
University of Delaware, Newark, DE 19716, USA
}




\pacs{12.60.Jv, 12.10.Dm, 14.80.Ly}

\begin{abstract}
We examine the low scale particle spectroscopy of an $SO(10)$ (or equivalently $SU(5)$) inspired supersymmetric model with 
non-universal gaugino masses. The model assumes minimal supergravity and contains the same number of fundamental parameters 
as the constrained minimal supersymmetric model (CMSSM.) Realistic solutions compatible with dark matter and other applicable 
experimental constraints are shown to exist for both positive and negative signs of the MSSM parameter $\mu$. 
We present several benchmark points which will be tested at the LHC and by the ongoing direct and indirect dark matter detection experiments.
\end{abstract}

\maketitle

\section{Introduction} 
The constrained minimal supersymmetric model (CMSSM), also referred to as mSugra, is based on the standard model gauge symmetry 
$SU(3)\times SU(2)\times U(1)$ and has the fewest number of fundamental parameters arising from the supersymmetric extension of the SM. 
Supersymmetry breaking in the CMSSM originates in some unspecified `hidden' sector, which is then transmitted through gravity to 
our `visible' sector. The lightest neutralino (LSP) in the CMSSM is stable and a leading dark matter candidate particle. Intense 
searches including direct, indirect and at the LHC, are currently underway to find the LSP. A flurry of recent papers by the ATLAS \cite{ATLAS-daCosta:2011qk} 
and CMS \cite{CMS} experiments at the LHC are beginning to constrain the  CMSSM parameter space, which translates into new lower bounds on a variety 
of sparticle masses including the gluino and the squarks of the first two families.

The apparent unification at $M_{G} \sim 3\times 10^{16}\ {\rm GeV}$ of the CMSSM gauge couplings strongly suggests the presence of an underlying 
grand unified theory (GUT) such as $SU(5)$ or $SO(10)$. Since the three MSSM gauge multiplets reside in a single unified gauge multiplet 
of $SU(5)$ or $SO(10)$, it would seem quite natural that the various MSSM gauginos all acquire the same universal mass $M_{1/2}$ at $M_{G}$. 
However, the universal gaugino mass assumption is, 
 in fact, not a general consequence of gravity mediated SUSY breaking. 
In the gravity mediation scenario, the gaugino masses are given by 
 non-zero $F$-term of the gauge kinetic function which, 
 generally, is in one of the irreducible representations of a symmetric product 
 of two adjoint representations and hence, not necessarily a singlet. 
Therefore, if such a non-singlet develops a non-zero $F$-term, 
 the MSSM gaugino masses can be expected to be non-universal. 
Interestingly, if we assume that a single non-zero $F$-term 
 dominates the gauge kinetic function, 
 the ratios between the MSSM gaugino masses are completely determined 
 by group theory, depending only on the way  
 the MSSM gauge multiplets are embedded in the symmetric product 
 of two adjoint representations of the GUT gauge group. 
This means that the number of free parameters in a setup with 
non-universal gaugino masses (NUGM) remains the same as in the CMSSM, at least in the gauge sector.

In the MSSM, the lightest neutralino, if it is the LSP, is 
 a primary candidate for the dark matter particle. 
Since the neutralino is an admixture of gauginos and Higgsinos, 
 its mass and interaction with other (s)particles are 
 determined by the masses of bino, wino and Higgsinos. 
Thus, a boundary condition involving non-universal gaugino masses 
 can dramatically alter the phenomenology of neutralino dark matter 
 from the CMSSM case \cite{NUG-DM}. 
In addition, the neutralino LSP plays a key role in the SUSY search 
 at high energy colliders, and composition and mass of 
 the neutralino being different from the CMSSM case has impact 
 on SUSY searches at the Tevatron \cite{NUG-Tevatron} 
 and at the Large Hadron Collider (LHC) \cite{NUG-LHC}.

In this paper, we investigate the  TeV scale particle spectroscopy of SUSY $SO(10)$  
and $SU(5)$ inspired model with non-universal gaugino masses. 
Taking a variety of phenomenological constraints into account, 
 in particular, the relic density of neutralino dark matter, 
 we identify the allowed regions of the input SSB parameters 
 for various $\tan \beta$ values. We consider both signs of the MSSM $\mu$-parameter. 
For the allowed parameter region, we calculate 
 the spin-independent (SI) and spin-dependent (SD) cross sections 
 for neutralino elastic scattering off a nucleon, 
 and compare our results with the current and proposed future bounds 
 from direct and indirect dark mater search experiments. 
We also present several benchmark points from the allowed region 
 and show the sparticle mass spectra, which can be tested 
 at the LHC. 
The mass spectra are compared with those from the CMSSM 
with suitably fixed parameter sets.

\section{Non-Universal Gaugino Masses from $SO(10)$ and $SU(5)$ 
\label{model}}

We first review how the non-universal gaugino masses arise 
 in the SO(10) GUT. 
In gravity mediation, we introduce a higher dimensional operator 
 between the gauge field strength superfield ${\cal W}^a$ and 
 a hidden sector chiral superfield $\Phi_{ab}$, 
\begin{eqnarray} 
 {\cal L}=\int d^2 \theta \frac{\Phi_{ab}}{M_P} {\cal W}^a {\cal W}^b, 
\end{eqnarray}
where $a,b=1,2, \cdots , 45$ are the group indices 
 for SO(10), and $M_P = 2.4 \times 10^{18}$ GeV 
 is the reduced Planck mass. 
In this paper, we consider only a single hidden sector field 
 $\Phi_{ab}$ whose non-zero $F$-term breaks SUSY and generates 
 the gaugino masses: 
\begin{eqnarray}
 {\cal L}= \int d^2 \theta \frac{\Phi_{ab}}{M_P} {\cal W}^a {\cal W}^b 
 \supset \frac{F_{\Phi_{ab}}}{M_P} \lambda^a \lambda^b.  
\end{eqnarray}

In SO(10), a possible representation of 
 the hidden sector field is given by one of the irreducible 
 representations contained in the symmetric product of 
 two adjoint 45-dimensional representations: 
\begin{eqnarray} 
( {\bf 45} \otimes {\bf 45})_{\rm sym} 
 = {\bf 1} \oplus {\bf 54} \oplus {\bf 210} \oplus {\bf 770}. 
\label{Eq3}
\end{eqnarray} 
If the hidden sector field is not the singlet, 
 non-universal masses for the MSSM gauginos are generated. 
However, the ratio  between the MSSM gaugino masses is determined  
by the embedding of the SM gauge groups 
with in each irreducible representation
 \cite{NUGmassformula1, NUGmassformula-Correct}.

Among the three possibilities in Eq.(\ref{Eq3}), we investigate $\Phi_{ab}$ in the {\bf 54} representation
 in this paper. 
In fact, this is the most reasonable case from the theoretical point of view. 
We are considering SUSY $SO(10)$ as a more fundamental 
 theory within which the MSSM is embedded. 
We may expect that the $SO(10)$ model is further unified 
 into a more fundamental theory including gravity, 
 most likely  some string theory at the Planck scale. 
Note that this picture constrains the field representations  
 introduced in the model, since large representations carry a large $\beta$-function, and this can cause 
the $SO(10)$ gauge coupling to blow up below the Planck scale. 
Indeed, the introduction of an irreducible representation larger 
 than 126 is excluded by this argument \cite{Darwin}. 
With the {\bf 54}-plet of hidden sector field, 
 the ratios of the non-universal gaugino masses is found to be 
 \cite{NUGmassformula-Correct} 
\begin{eqnarray}
 M_1 : M_2 : M_3 = - \frac{1}{2} : -\frac{3}{2} : 1 
\label{NUGratio}
\end{eqnarray}

The above discussion is readily extended to $SU(5)$  
, and the gluino mass ratios turn out to be the same as in Eq.(\ref{NUGratio}) for a hidden sector
field in the 24-dimensional adjoint 
 representation \cite{NUGmassformula-Correct}. 
Thus, our analysis in this paper applies both to $SO(10)$ and $SU(5)$ models. 
However, note that in the $SU(5)$ GUT, the matter multiplets of each family are not 
 completely unified in a single representation and in general, 
the  masses of sfermions in {\bf 5}$^*$ and {\bf 10} representations 
 can be non-universal. 
This non-universality of sfermion masses can serve as a probe 
 to discriminate the underlying GUT gauge groups \cite{SU5probe}.

In the following, we identify the parameter region which 
 is consistent with a variety of phenomenological constraints. 
Since the ratio between the non-universal gaugino masses is fixed  
 as in Eq.~(\ref{NUGratio}), the number of free parameters 
 remains the same as in the CMSSM. 
We use the notation $M_3=M_{1/2}$, so that 
 $M_1=-\frac{1}{2} M_{1/2}$ and $M_{2}=- \frac{3}{2} M_{1/2}$. 
The fundamental parameters of the {\bf $SO(10)/SU(5)$} model that we consider
 are as follows: 
\begin{eqnarray} 
  m_0, M_{1/2}, A_0, \tan \beta, {\rm sign}(\mu).  
 \label{parameters} 
\end{eqnarray}

\section{Phenomenological constraints and scanning procedure
\label{constraintsSection}}

We employ the ISAJET~7.80 package~\cite{ISAJET}  to perform random 
 scans over the parameter space listed in Eq.~(\ref{parameters}). 
In this package, the weak scale values of gauge and third 
 generation Yukawa couplings are evolved to 
 $M_{\rm GUT}$ via the MSSM renormalization group equations (RGEs) 
 in the $\overline{DR}$ regularization scheme. 
We do not strictly enforce the unification condition 
 $g_3=g_1=g_2$ at $M_{\rm GUT}$, since a few percent deviation 
 from unification can be assigned to unknown GUT-scale threshold 
 corrections~\cite{Hisano:1992jj}. 
At $M_{\rm GUT}$, the boundary conditions given in 
 Eq.~(\ref{parameters}) are imposed and all the SSB parameters, 
 along with the gauge and Yukawa couplings, are evolved
 back to the weak scale $M_{\rm Z}$. 
The impact of the neutrino Dirac Yukawa coupling 
 in the running of the RGEs is significant only 
 for relatively large values ($\sim 2$ or so) \cite{Gomez:2009yc}. 
In the SO(10) GUT we expect the largest Dirac coupling 
 to be comparable to the top Yukawa coupling 
 ($\sim 0.5$ at $M_{\rm GUT}$) and thus we safely neglect 
 effects of the neutrino Dirac Yukawa coupling 
 in our analysis.

In the evaluation of Yukawa couplings the SUSY threshold 
 corrections~\cite{Pierce:1996zz} are taken into account 
 at the common scale $M_{\rm SUSY}= \sqrt{m_{\tst_L}m_{\tst_R}}$. 
The entire parameter set is iteratively run between 
 $M_{\rm Z}$ and $M_{\rm GUT}$ using the full 2-loop RGEs 
 until a stable solution is obtained. 
To better account for leading-log corrections, one-loop step-beta 
 functions are adopted for gauge and Yukawa couplings, and 
 the SSB parameters $m_i$ are extracted from RGEs at multiple scales
 $m_i=m_i(m_i)$. 
The RGE-improved 1-loop effective potential is minimized 
 at an optimized scale $M_{\rm SUSY}$, which effectively 
 accounts for the leading 2-loop corrections. 
Full 1-loop radiative corrections are incorporated 
 for all sparticle masses.

The requirement of radiative electroweak symmetry breaking 
 (REWSB)~\cite{Ibanez:1982fr} puts an important theoretical
 constraint on the parameter space. 
Another important constraint comes from limits on the cosmological 
 abundance of stable charged particles~\cite{Nakamura:2010zzi}. 
This excludes regions in the parameter space where charged 
 SUSY particles, such as $\ttau_1$ or $\tst_1$, 
 become the LSP. 
We accept only those solutions for which one of the neutralinos 
 is the LSP and saturates the dark matter relic abundance bound 
 observed by the Wilkinson Microwave Anisotropy Probe (WMAP). 

We perform random scans for the following parameter range:
\begin{align}
0\leq  m_{0} \leq 5\, \rm{TeV} \nonumber \\
0\leq M_{1/2}  \leq 2\, \rm{TeV} \nonumber \\
 \tan\beta = 10, 30, 50 \nonumber \\
 A_{0} = 0, 1,-1, 5,-5\, \rm{TeV}\nonumber \\
\mu < 0 ,  \mu > 0
\label{parameterRange}
\end{align}
 with $m_t = 173.1\, {\rm GeV}$ \cite{:2009ec}. 
The results are not too sensitive to one or two sigma variation 
 in the value of $m_t$. 
We use $m_b(m_Z)=2.83$ GeV which is hard-coded into ISAJET.

In scanning the parameter space, we employ the Metropolis-Hastings 
 algorithm as described in \cite{Belanger:2009ti}. 
All of the collected data points satisfy the requirement of REWSB,
 with the neutralino in each case being the LSP. 
Furthermore, all of these points satisfy the constraint 
 $\Omega_{\rm CDM}h^2 \le 10$. 
This is done so as to collect more points with a WMAP compatible 
 value of cold dark matter (CDM) relic abundance. 
For the Metropolis-Hastings algorithm, we only use 
 the value of $\Omega_{\rm CDM}h^2$ to bias our search. 
Our purpose in using the Metropolis-Hastings algorithm is 
 to be able to search around regions of acceptable 
 $\Omega_{\rm CDM}h^2$ more fully. 
After collecting the data, we impose the mass bounds on 
 all the particles~\cite{Nakamura:2010zzi} and use the
 IsaTools package~\cite{Baer:2002fv} to implement 
 the following phenomenological constraints:
\begin{table}[h]\centering
\begin{tabular}{rlc}
$m_h~{\rm (lightest~Higgs~mass)} $&$ \geq\, 114.4~{\rm GeV}$                    &  \cite{Schael:2006cr} \\
$BR(B_s \rightarrow \mu^+ \mu^-) $&$ <\, 5.8 \times 10^{-8}$                     &   \cite{:2007kv}      \\
$2.85 \times 10^{-4} \leq BR(b \rightarrow s \gamma) $&$ \leq\, 4.24 \times 10^{-4} \; (2\sigma)$ &   \cite{Barberio:2007cr}  \\
$0.53\le \frac {BR{(B_u\to\tau\nu_\tau)_{\rm MSSM} }}{{BR(B_u\to\tau\nu_\tau)_{\rm SM}}} $&$ \le 2.03 \; (2\sigma)$ & \cite{Eriksson:2008cx}\\
$\Omega_{\rm CDM}h^2 $&$ =\, 0.111^{+0.028}_{-0.037} \;(5\sigma)$               &  \cite{Komatsu:2008hk}    \\
$3.4 \times 10^{-10}\leq \Delta (g-2)_{\mu}/2 $&$ \leq\, 55.6 \times 10^{-10}~ \; (3\sigma)$ &  \cite{Bennett:2006fi} \\
 
\end{tabular}

\end{table}

\noindent 
We apply the experimental constraints successively on the data that
we acquire from ISAJET.

\section{Results\label{results}}

\begin{figure}
\centering
\subfiguretopcaptrue
\subfigure[\hspace {1mm} $A_0=0,\tan\beta=10,\mu>0$]{
\includegraphics[width=8cm]{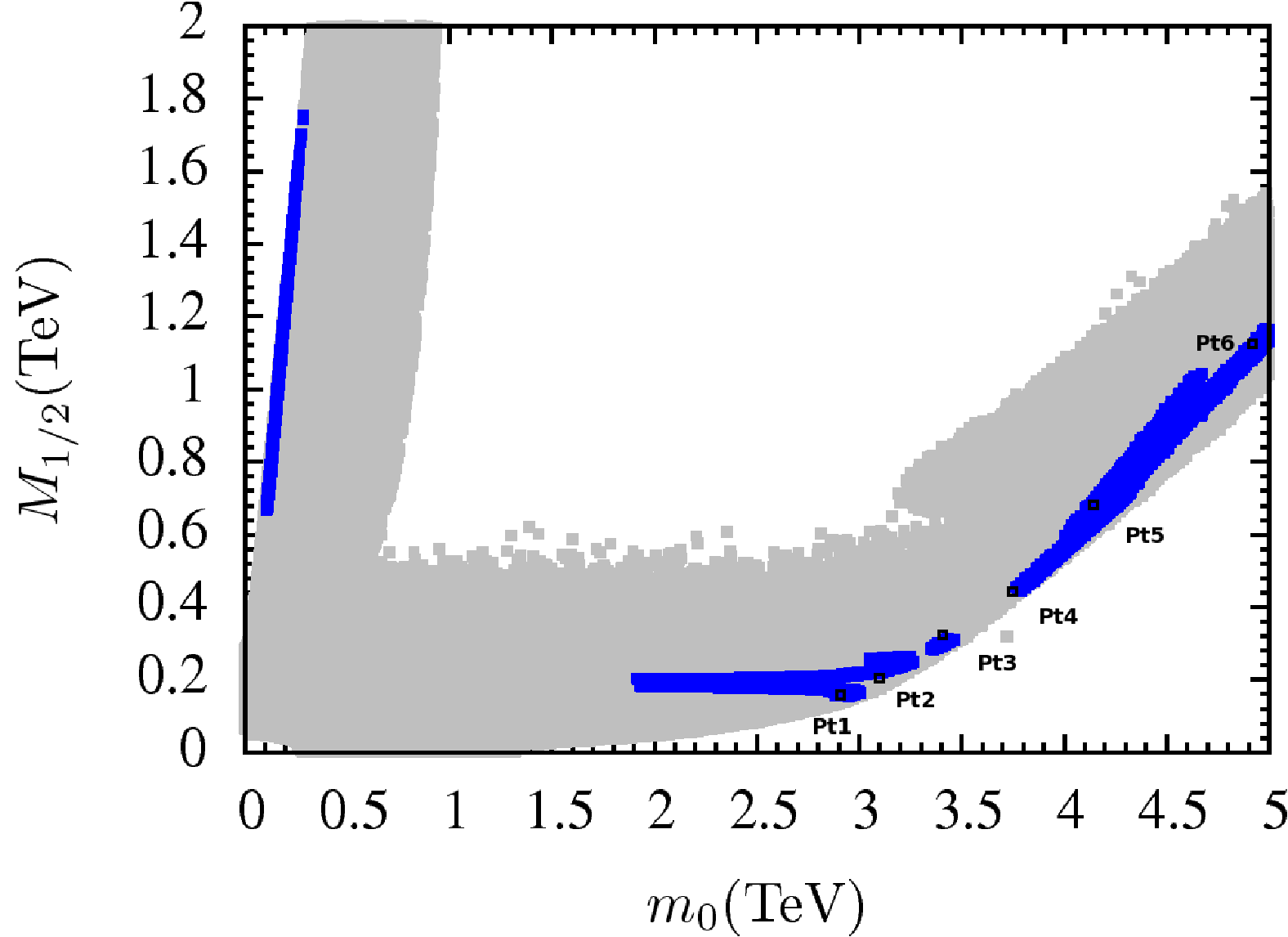}
}
\subfigure[\hspace {1mm} $A_0=0,\tan\beta=10,\mu<0$]{
\includegraphics[width=8cm]{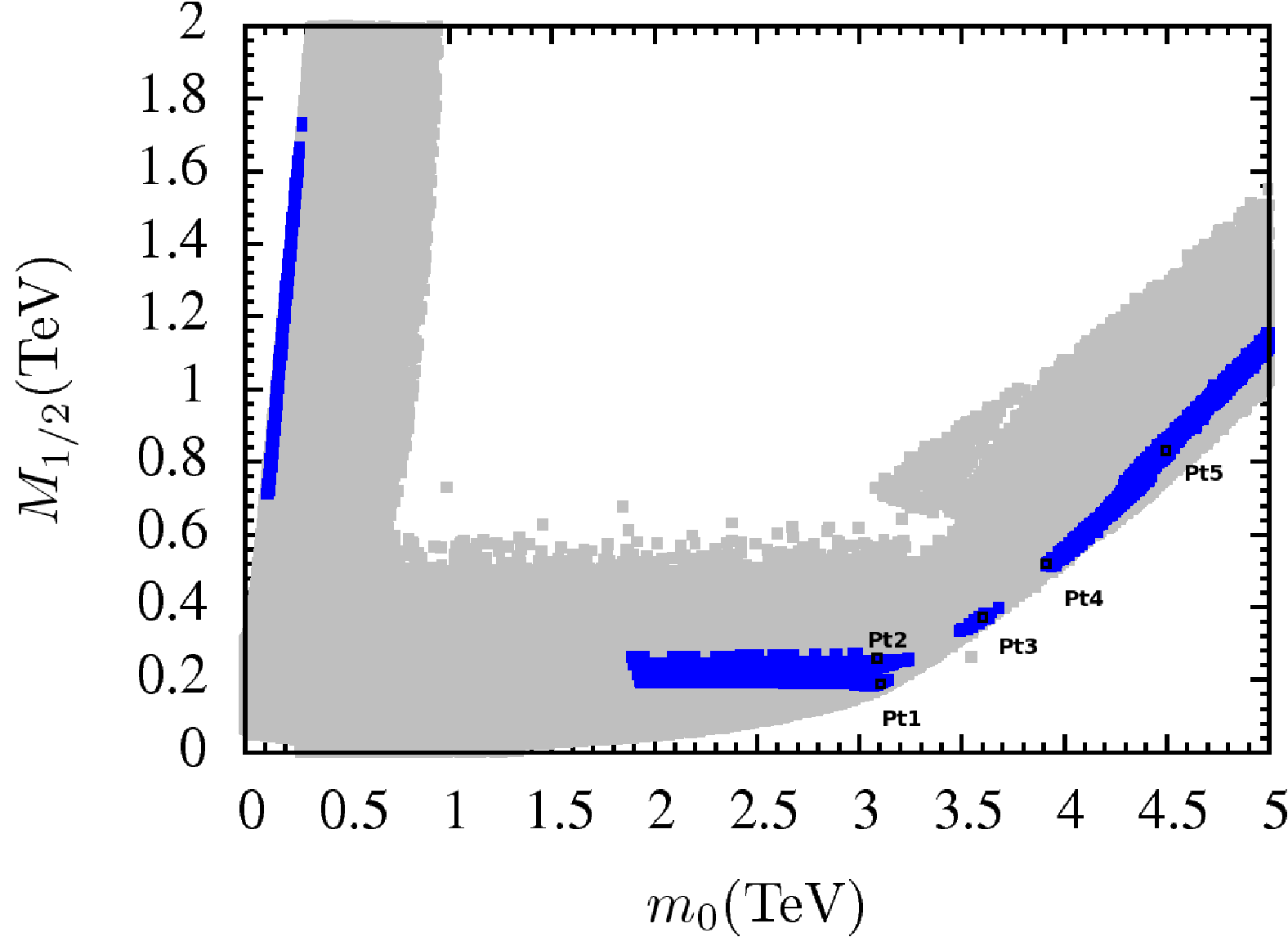}
}
\caption{
Plots in the ($M_{1/2}$, $m_0$) plane. 
Gray points are consistent with REWSB and $\tilde{\chi}^0_{1}$ LSP. 
Blue points satisfy the WMAP bounds on $\tilde{\chi}^0_1$ 
dark matter abundance, particle mass bounds, 
constraints from $BR(B_s\rightarrow \mu^+ \mu^-)$ and 
$BR(b\rightarrow s \gamma)$ and {\bf $BR(B_u\to\tau\nu_\tau)$}. 
Approximate locations of benchmark points 
listed in Tables~\ref{table1} and \ref{table2} are also shown. 
}
\label{m0M12A0TB10}
\end{figure}

Figure~\ref{m0M12A0TB10} shows the results 
 in the $(M_{1/2}$, $m_0$) plane for $\tan \beta=10$, 
 $A_0 = 0$  and ${\rm sign}(\mu)=\pm$.  
Gray points are consistent with successful REWSB and 
 the requirement of neutralino LSP. 
Blue points satisfy the WMAP bounds on neutralino dark matter 
 abundance, particle mass bounds, 
as well as constraints from $BR(B_s\rightarrow \mu^+ \mu^-)$,
 $BR(b\rightarrow s \gamma)$ and $BR(B_u\to\tau\nu_\tau)$.
In Figure~\ref{m0M12A0TB10}, the small boxes show benchmark points 
 for each of which the particle mass spectra are listed 
 in Tables~\ref{table1} and \ref{table2}. 
We have chosen these benchmark points 
 from regions in Figure~\ref{SIA0TB10} and \ref{SDA0TB10}, 
 which can be explored by future dark matter search experiments 
 (see below).

A variety of experiments are underway to directly detect dark matter
 particles through their elastic scatterings off nuclei. 
The most stringent limits on the (spin-independent) elastic 
 scattering cross section have been reported by 
 the recent CDMS-II and XENON100 experiments. 
In Figure~\ref{SIA0TB10} (the color coding is the same 
 as in Figure~\ref{m0M12A0TB10}) we show the results for the spin-independent 
 elastic scattering cross section along with the current upper bounds 
 by the CDMS-II \cite{CDMSII} (solid black line) and XENON100 \cite{Aprile} (solid red line)
 experiments, as well as the future reach of 
 the SuperCDMS(SNOLAB) \cite{Bruch}(dotted black line ) and XENON1T \cite{XENON1T} (dotted red line). 
In the ($\sigma_{\rm SI}, m_{\tilde{\chi}_1^{0}}$) plane, 
 there are dips in the resultant cross sections. 
Two of them around $m_{\tilde{\chi}_1^{0}} \sim  45$ GeV 
and $m_{\tilde{\chi}_1^{0}} \sim 57$ GeV 
  correspond to Z- and Higgs resonances, respectively. 
Around these parameter regions, 
the neutralino annihilation cross sections are enhanced 
 by the resonances and as a result, the correct relic abundance 
 can be achieved with relatively  small coupling constants. 
Thus, the corresponding spin-independent cross sections 
 are reduced due to the small coupling constants. 
Another dip around $m_{\tilde{\chi}_1^{0}} \sim 200$ GeV 
 in Figure~\ref{SIA0TB10} (a) corresponds 
 to a cancellation between effective couplings 
 of neutralino with up-quark and down-quark in Higgs exchange processes. 
As has been reported in \cite{Ellis:2000ds}, 
 this cancellation occurs when the relative signs 
 between $M_{1,2}$ and $\mu$ are opposite. 
Recall that in our convention, $M_3 > 0$ and $M_{1,2} < 0$, 
 so that this cancellation occurs for $\mu >0$. 
The small boxes show the benchmark points corresponding 
 to those listed in Tables~\ref{table1} and \ref{table2}. 
These benchmark points have been chosen with the criterion that 
 they should lie between the present experimental limits 
 and future reaches. 
In the region from which the benchmark points are chosen, 
 the neutralino LSP has sizable Higgsino components 
 which, in turn, enhance the spin-independent 
 neutralino-nucleon cross sections.  
All of the benchmark points can be tested by XENON1T, 
 while SuperCDMS can explore some of them.

Neutralino dark matters in the galactic halo may become 
 gravitationally trapped in the Sun and accumulate 
 in its center, where they can annihilate each other
 and produce high energy neutrinos. 
Since neutrinos can escape and reach the Earth, 
 the neutralino annihilations can be indirectly 
 detected by observing an excess of such high energy neutrinos 
 from the Sun. 
The most stringent limits on the neutrino flux from the Sun 
 have been reported by Super-Kamiokande \cite{SK} 
 and IceCube \cite{IceCube} experiments, which provide 
 the upper limit on the spin-dependent elastic scattering 
 cross section of neutralino dark matter off a nucleon. 
The spin-dependent scattering cross sections 
 along with the current upper bounds 
 and future reach are depicted in Figure~\ref{SDA0TB10}. 
In this figure too, 
 the color coding is the same as in Figure~\ref{m0M12A0TB10}. 
In the ($\sigma_{\rm SD}, m_{\tilde{\chi}_1^{0}}$) plane 
 we show the results for the spin-dependent elastic scattering cross section
 along with the current bounds from Super-Kamiokande (black line)
 and IceCube (dotted black line) experiments, 
 together with the future reach of IceCuce DeepCore experiment (dotted red line). 
The small boxes show the approximate locations of benchmark points
 corresponding to Tables~\ref{table1} and \ref{table2}.
For these points, the sizable Higgsino components 
 of the neutralino LSP enhances the scattering cross section. 
All benchmark points except for point 1 are testable 
 by the future IceCube DeepCore experiment. 
Although the benchmark point 1 has a neutralino mass below 
 the energy threshold of the IceCuce DeepCore experiment, 
 such a light neutralino can provide characteristic signatures 
  in collider experiments.

\begin{figure}
\centering
\subfiguretopcaptrue
\subfigure[\hspace {1mm} $A_0=0,\tan\beta=30,\mu>0$] {
\includegraphics[width=8cm]{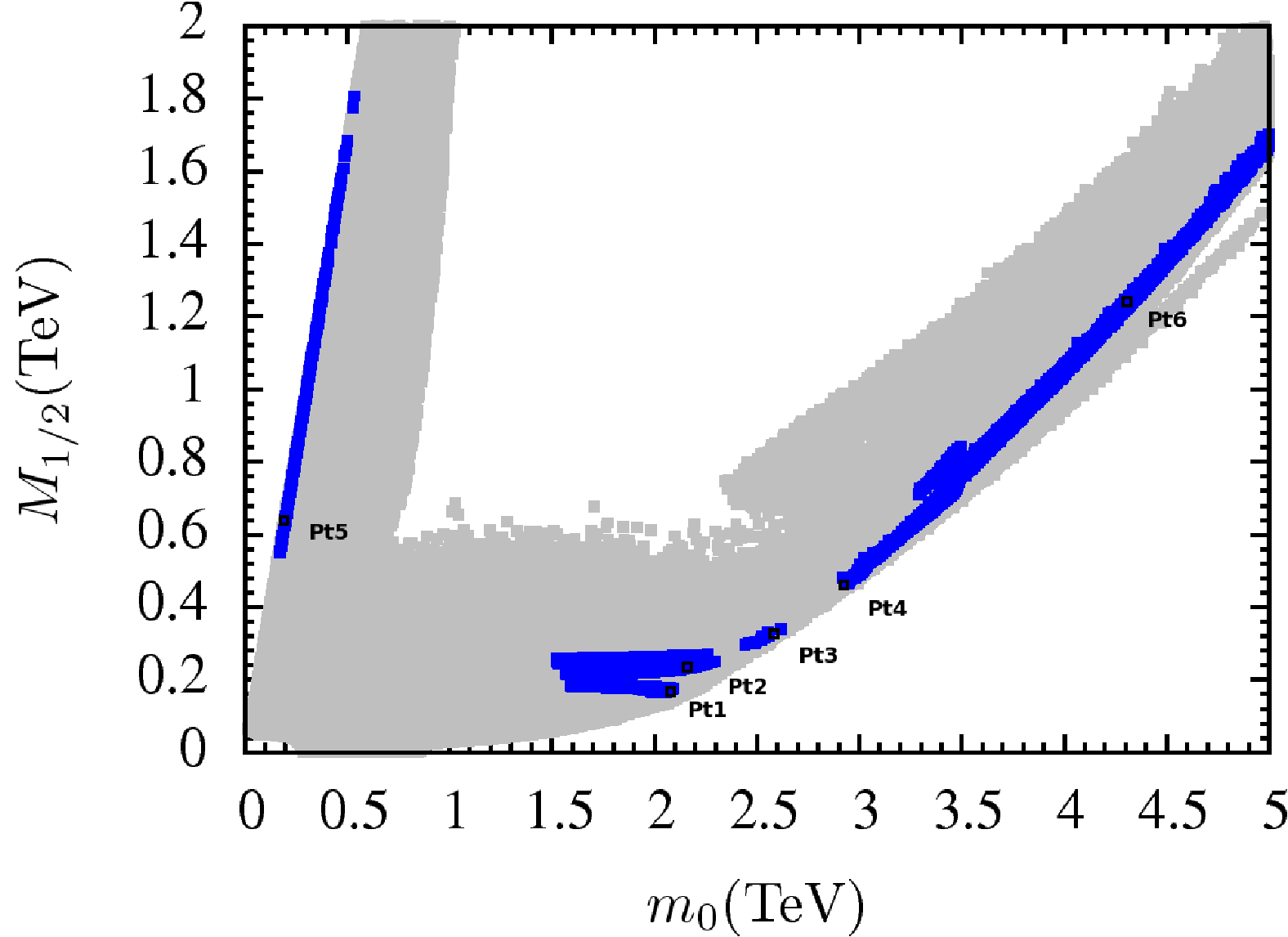}
} \subfigure[\hspace {1mm} $A_0=0,\tan\beta=30,\mu<0$]{
\includegraphics[width=8cm]{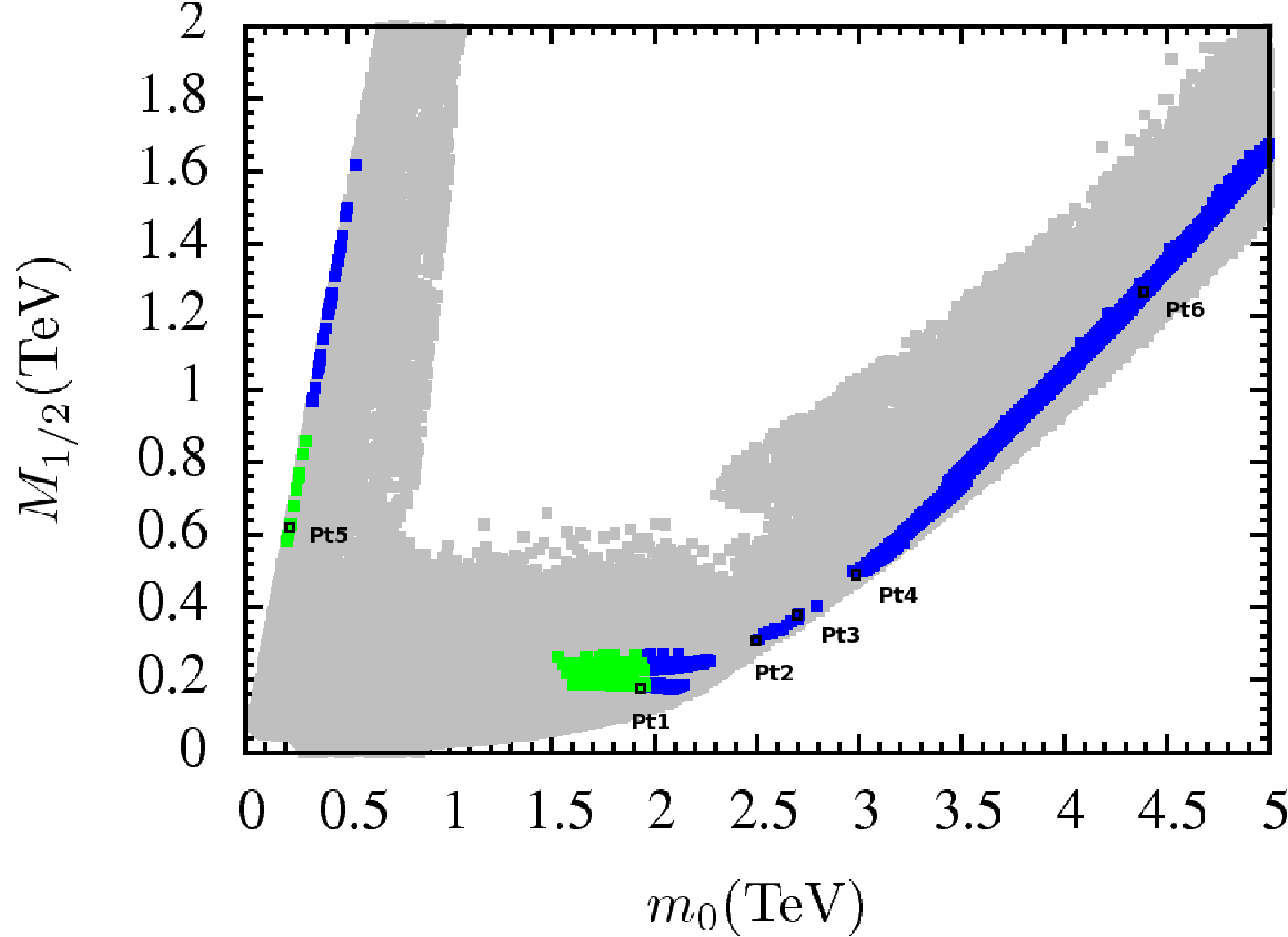}
}
\caption{
Plots in the ($M_{1/2}, m_0$) plane. 
Gray points are consistent with REWSB and $\tilde{\chi}^0_{1}$ LSP. 
Blue points satisfy the WMAP bounds on $\tilde{\chi}^0_1$ 
 dark matter abundance, particle mass bounds, 
 constraints from $BR(B_s\rightarrow \mu^+ \mu^-)$,
 $BR(b\rightarrow s \gamma)$ and $BR(B_u\to\tau\nu_\tau)$.
Green points belong to the subset of blue points that satisfies 
 all constraints including $\Delta(g-2)_\mu/2$.
Approximate locations of benchmark points 
 listed in Tables~\ref{table3}
and \ref{table4} are also shown.}
\label{m0M12A0TB30}
\end{figure}

\begin{figure}
\centering
\subfiguretopcaptrue
\subfigure[\hspace {1mm} $A_0=0,\tan\beta=30,\mu>0$]{
\includegraphics[width=8cm]{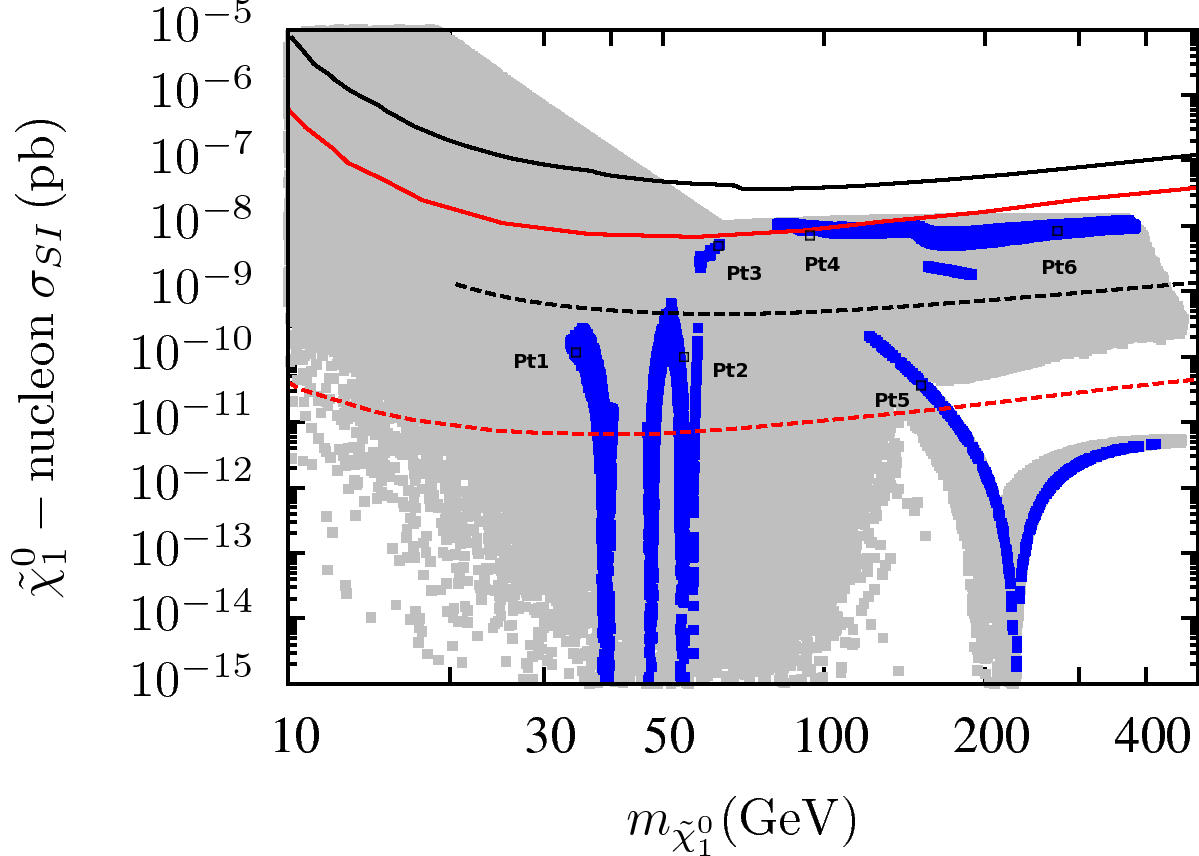}
} \subfigure[\hspace {1mm} $A_0=0,\tan\beta=30,\mu<0$]{
\includegraphics[width=8cm]{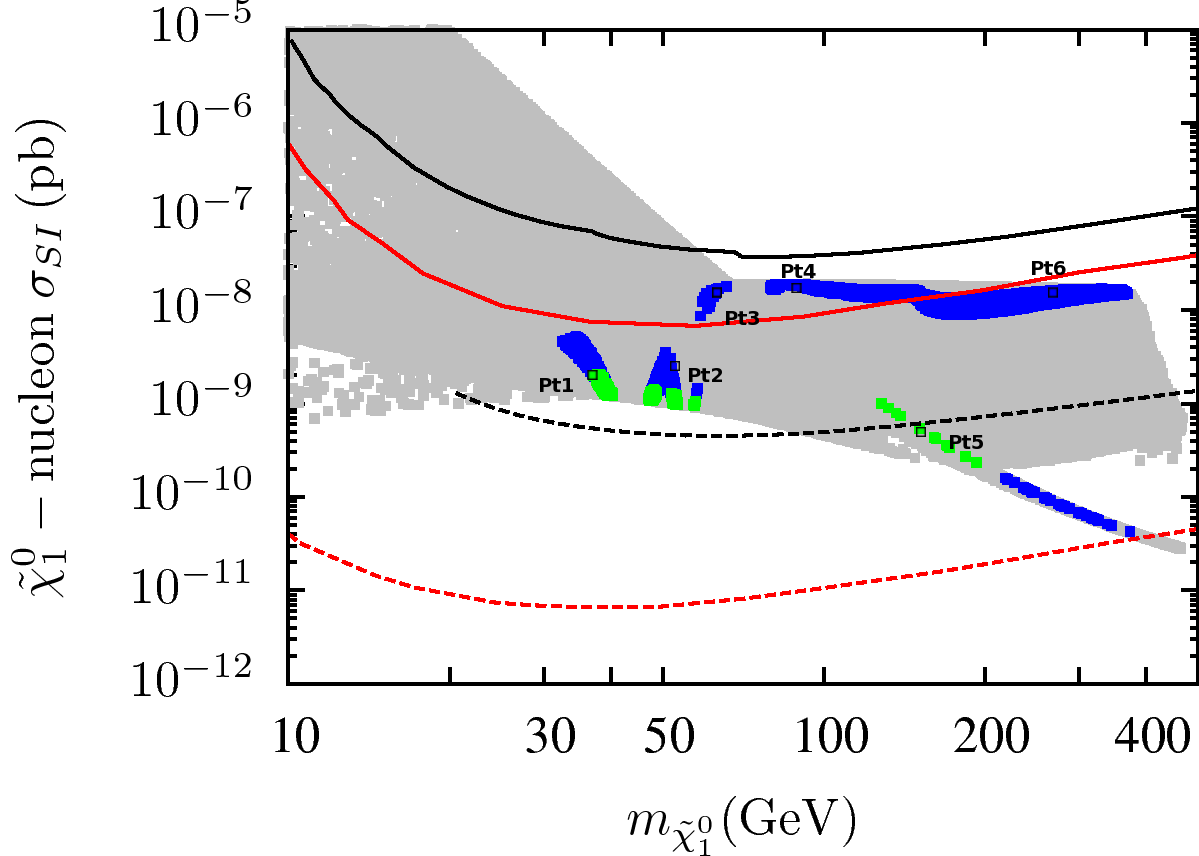}
}
\caption{
Spin-independent elastic scattering cross section 
 of neutralino dark matter 
 in the ($\sigma_{\rm SI}, m_{\tilde{\chi}_1^{0}}$) plane. 
Color coding is the same as in Figure~\ref{m0M12A0TB30}.
Current upper bounds 
 from CDMS-II (XENON100) 
 are depicted as solid black (red)lines.
Future reach of SuperCDMS(SNOLAB) (dotted black line) 
 and XENON1T (dotted red line) are shown.  
Approximate locations of benchmark points 
listed in Tables~\ref{table3} and \ref{table4} are also shown. 
} 
\label{SIA0TB30}
\end{figure}

\begin{figure}
\centering
\subfiguretopcaptrue
\subfigure[\hspace {1mm} $A_0=0,\tan\beta=30,\mu>0$]{
\includegraphics[width=8cm]{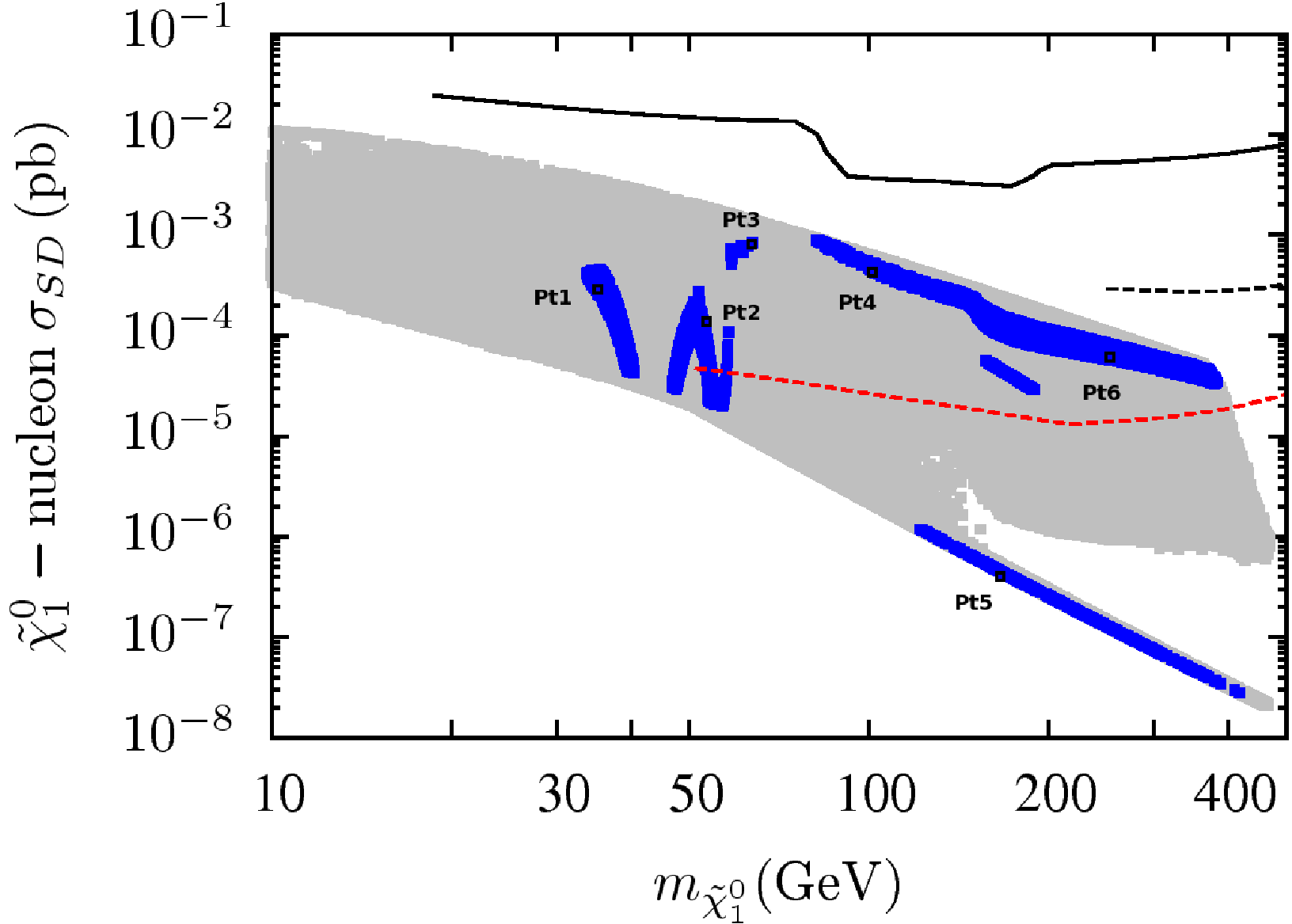}
} \subfigure[\hspace {1mm} $A_0=0,\tan\beta=30,\mu<0$] {
\includegraphics[width=8cm]{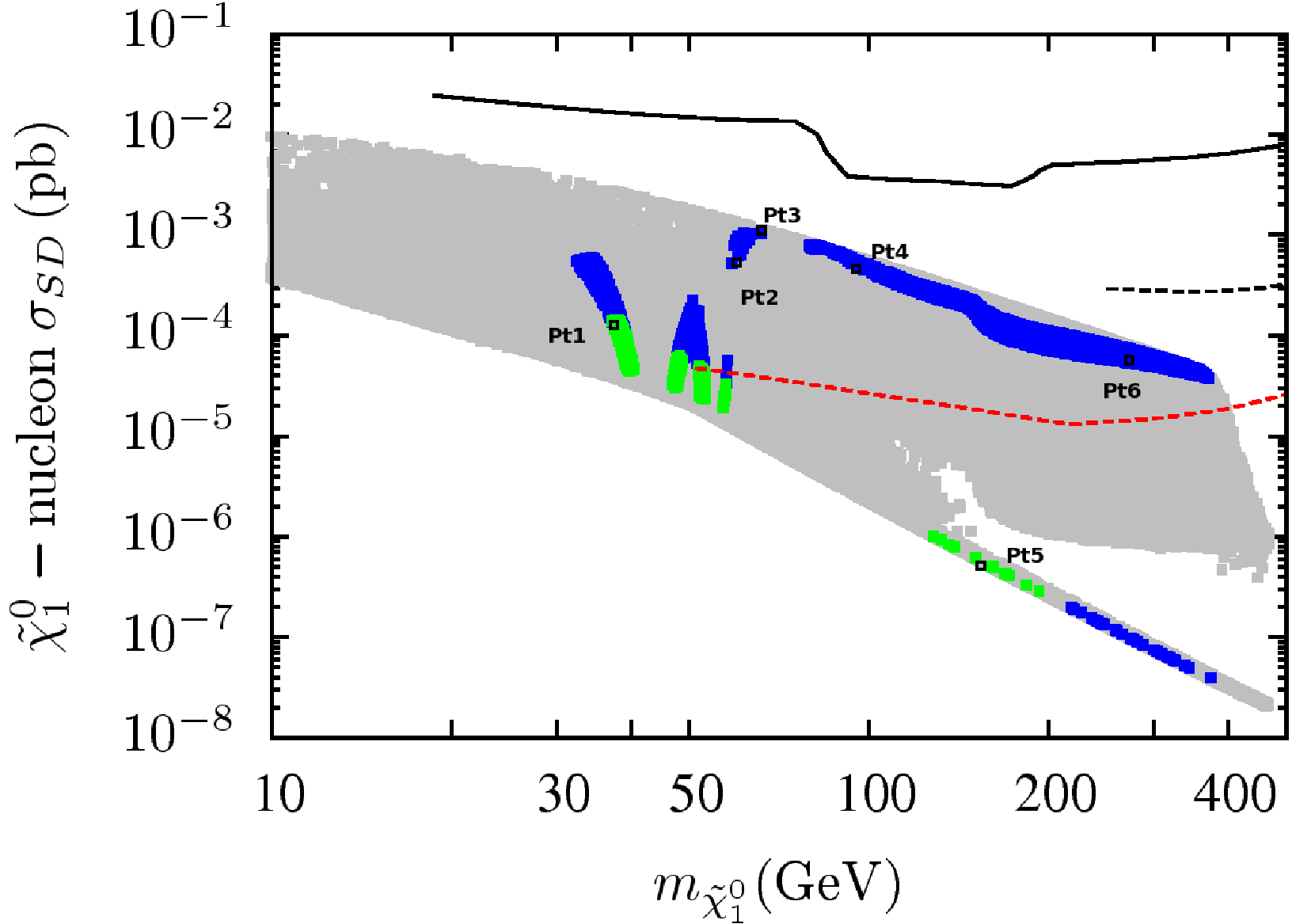}
}
\caption{
Spin-dependent elastic scattering cross section 
 of neutralino dark matter 
 in the ($\sigma_{\rm SD}, m_{\tilde{\chi}_1^{0}}$) plane. 
Color coding is the same as in Figure~\ref{m0M12A0TB30}.
Current upper bounds from Super-Kamiokande (black line), IceCube (dotted black line), and future reach of 
 IceCube DeepCore (dotted red line) are shown. 
Approximate locations of benchmark points 
listed in Tables~\ref{table3} and \ref{table4} are also shown. 
}
\label{SDA0TB30}
\end{figure}

\begin{figure}
\centering
\subfiguretopcaptrue
\subfigure[\hspace {1mm} $A_0=0,\tan\beta=50,\mu>0$]{
\includegraphics[width=8cm]{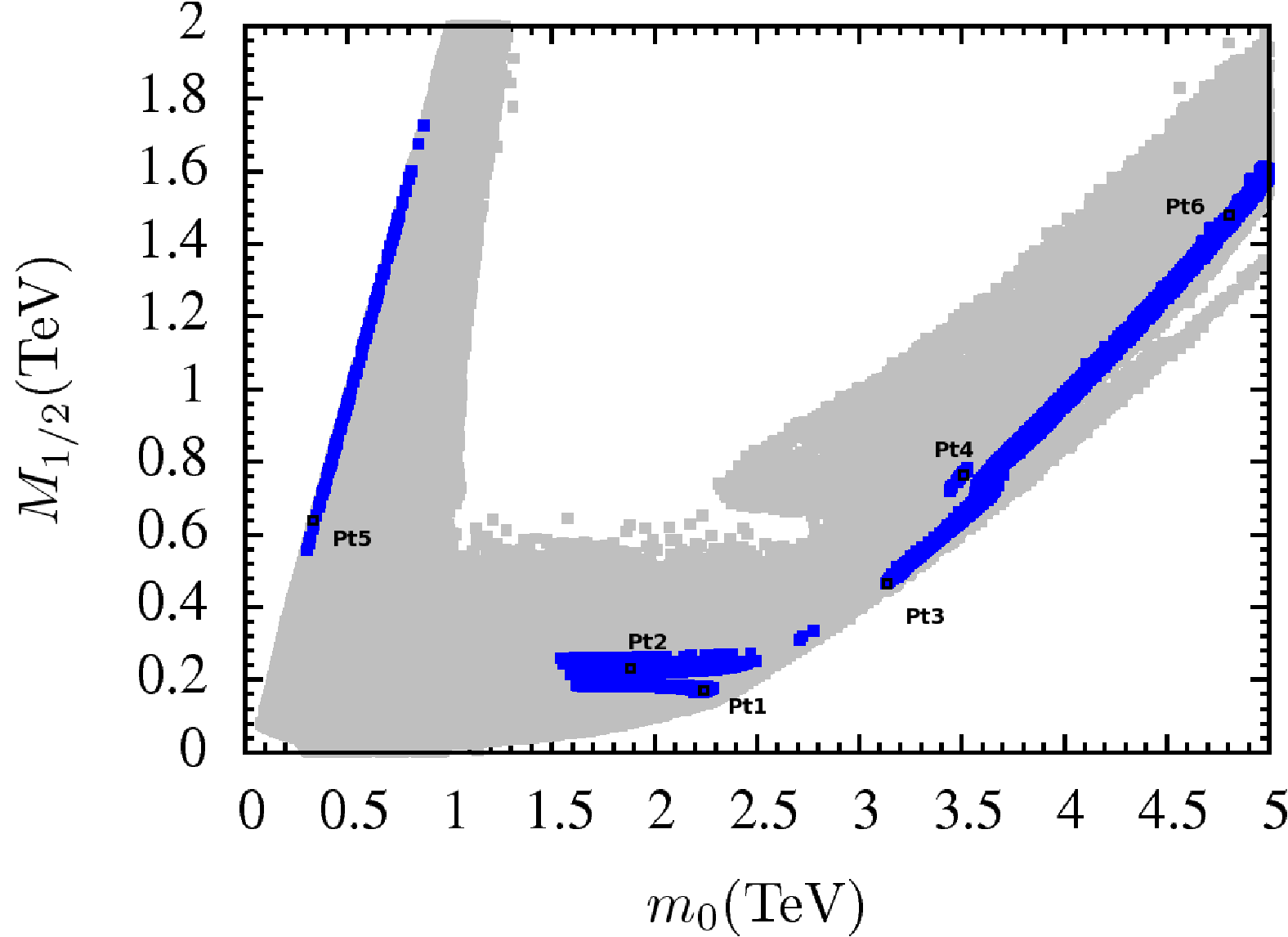}
} \subfigure[\hspace {1mm} $A_0=0,\tan\beta=50,\mu<0$]{
\includegraphics[width=8cm]{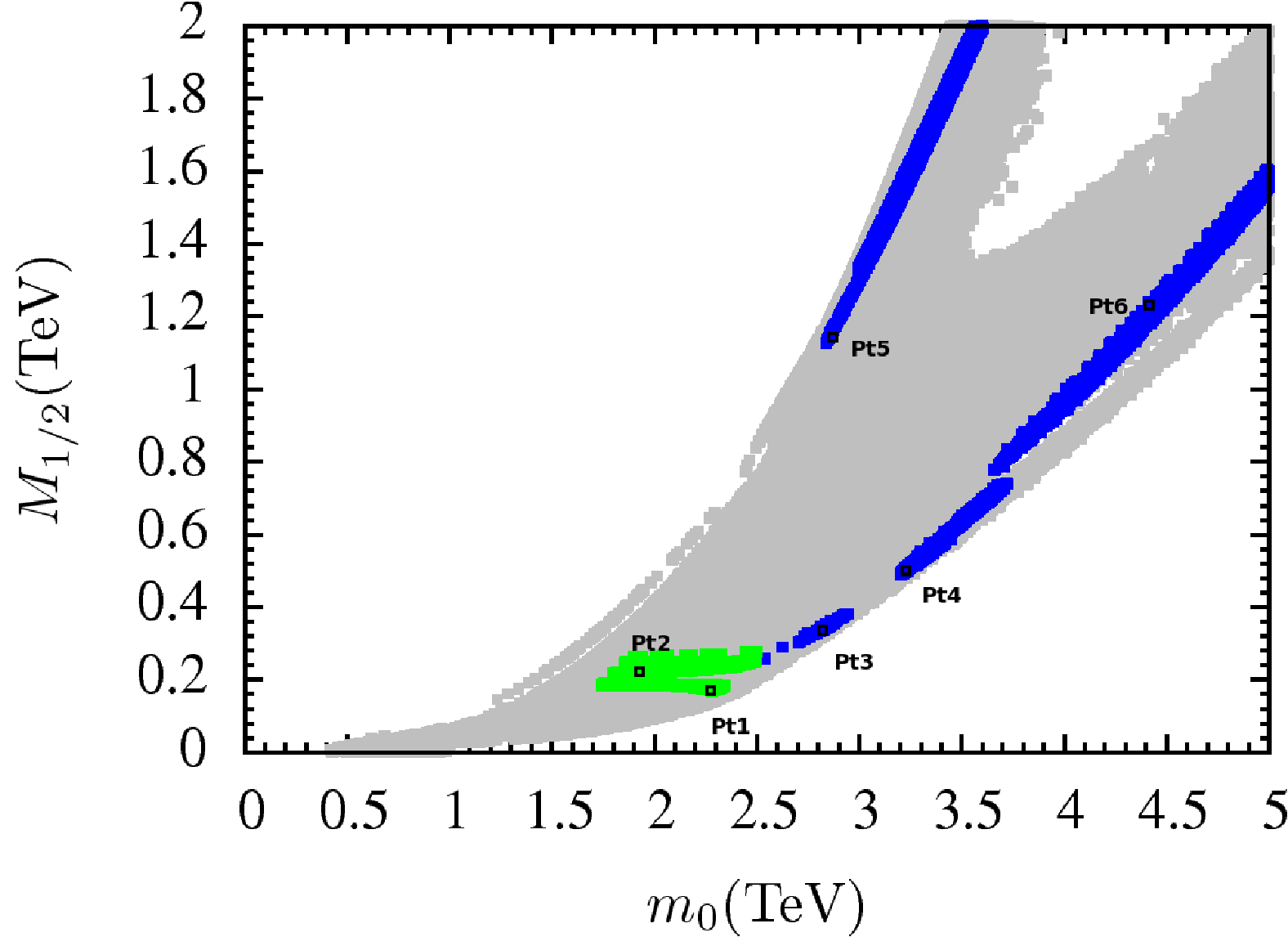}
}
\caption{
Plots in the ($M_{1/2}, m_0$) plane.
Gray points are consistent with REWSB and $\tilde{\chi}^0_{1}$ LSP.
Blue points satisfy the WMAP bounds on $\tilde{\chi}^0_1$
 dark matter abundance, particle mass bounds,
 constraints from $BR(B_s\rightarrow \mu^+ \mu^-)$,
 $BR(b\rightarrow s \gamma)$ and $BR(B_u\to\tau\nu_\tau)$.
Green points belong to the subset of blue points that satisfies
 all constraints including $\Delta(g-2)_\mu/2$.
Approximate locations of benchmark points 
 listed in Tables~\ref{table5} and \ref{table6} are also shown.
}
\label{m0M12A0TB50}
\end{figure}

\begin{figure}
\centering
\subfiguretopcaptrue
\subfigure[\hspace {1mm} $A_0=0,\tan\beta=50,\mu>0$]{
\includegraphics[width=8cm]{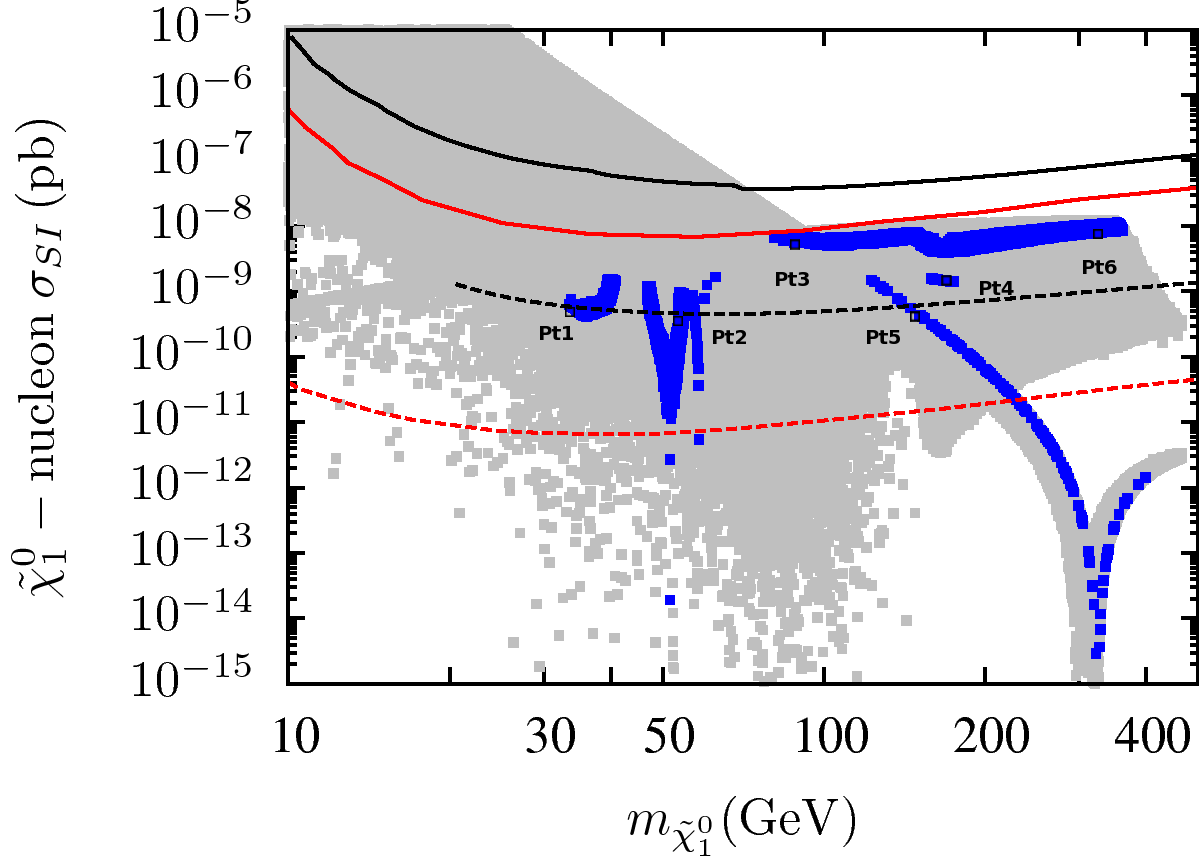}
} \subfigure[\hspace {1mm} $A_0=0,\tan\beta=50,\mu<0$]{
\includegraphics[width=8cm]{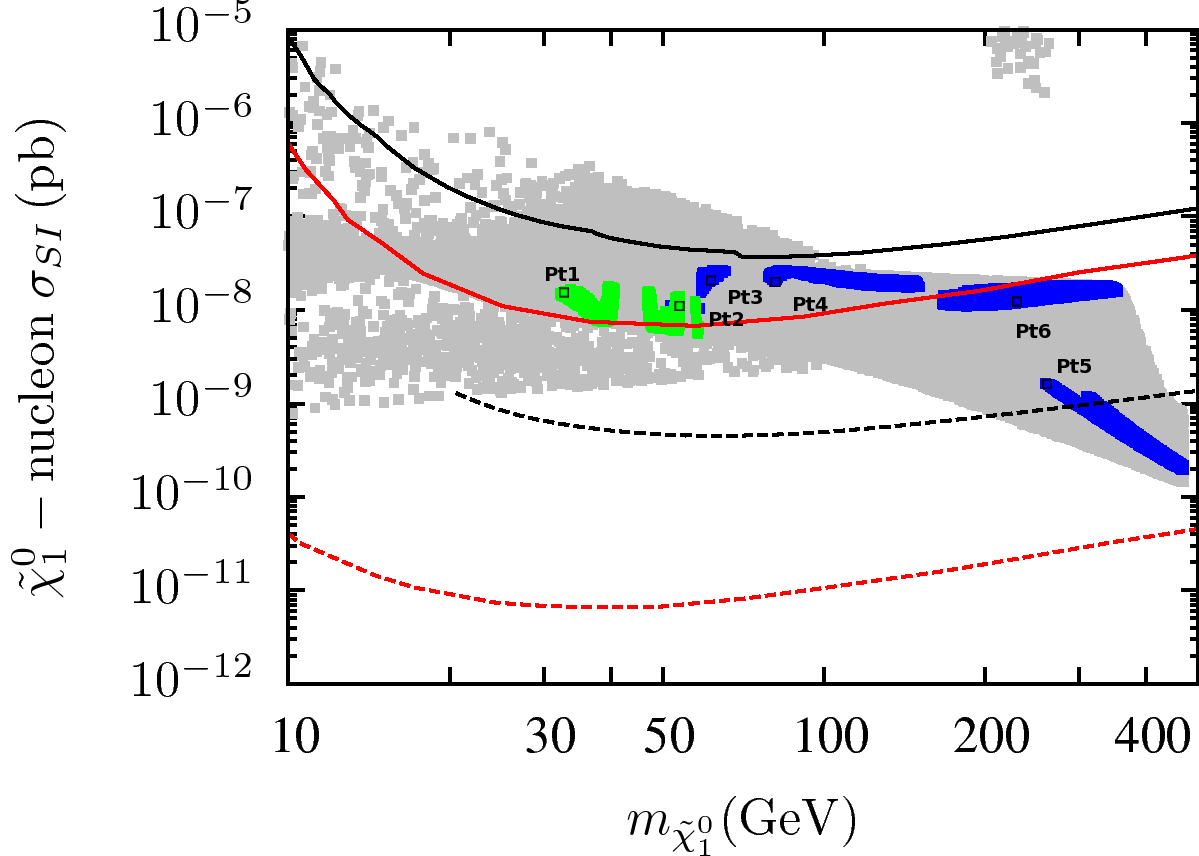}
}
\caption{
Spin-independent elastic scattering cross section 
 of neutralino dark matter 
 in the ($\sigma_{\rm SI}, m_{\tilde{\chi}_1^{0}}$) plane. 
Color coding is same as in Figure~\ref{m0M12A0TB30}. 
Current upper bounds 
 from the CDMS-II (XENON100) are depicted as solid black (red) lines. 
 Future reach of SuperCDMS(SNOLAB) (dotted black line) 
 and XENON1T (dotted red line) are shown.  
Approximate locations of benchmark points 
listed in Tables~\ref{table5} and \ref{table6} are also shown. 
}
\label{SIA0TB50}
\end{figure}

\begin{figure}
\centering
\subfiguretopcaptrue
\subfigure[\hspace {1mm} $A_0=0,\tan\beta=50,\mu>0$]{
\includegraphics[width=8cm]{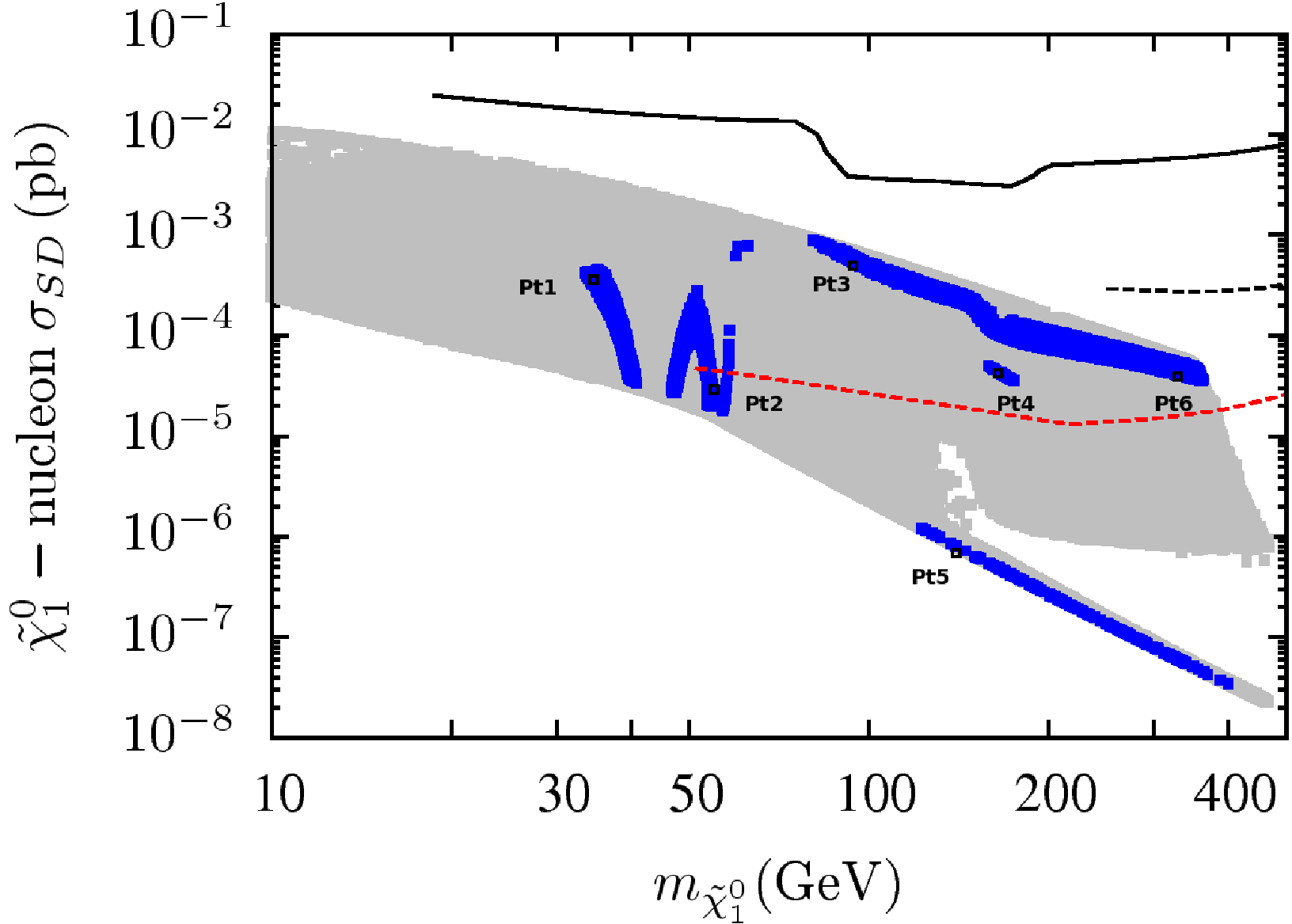}
} \subfigure[\hspace {1mm} $A_0=0,\tan\beta=50,\mu<0$]{
\includegraphics[width=8cm]{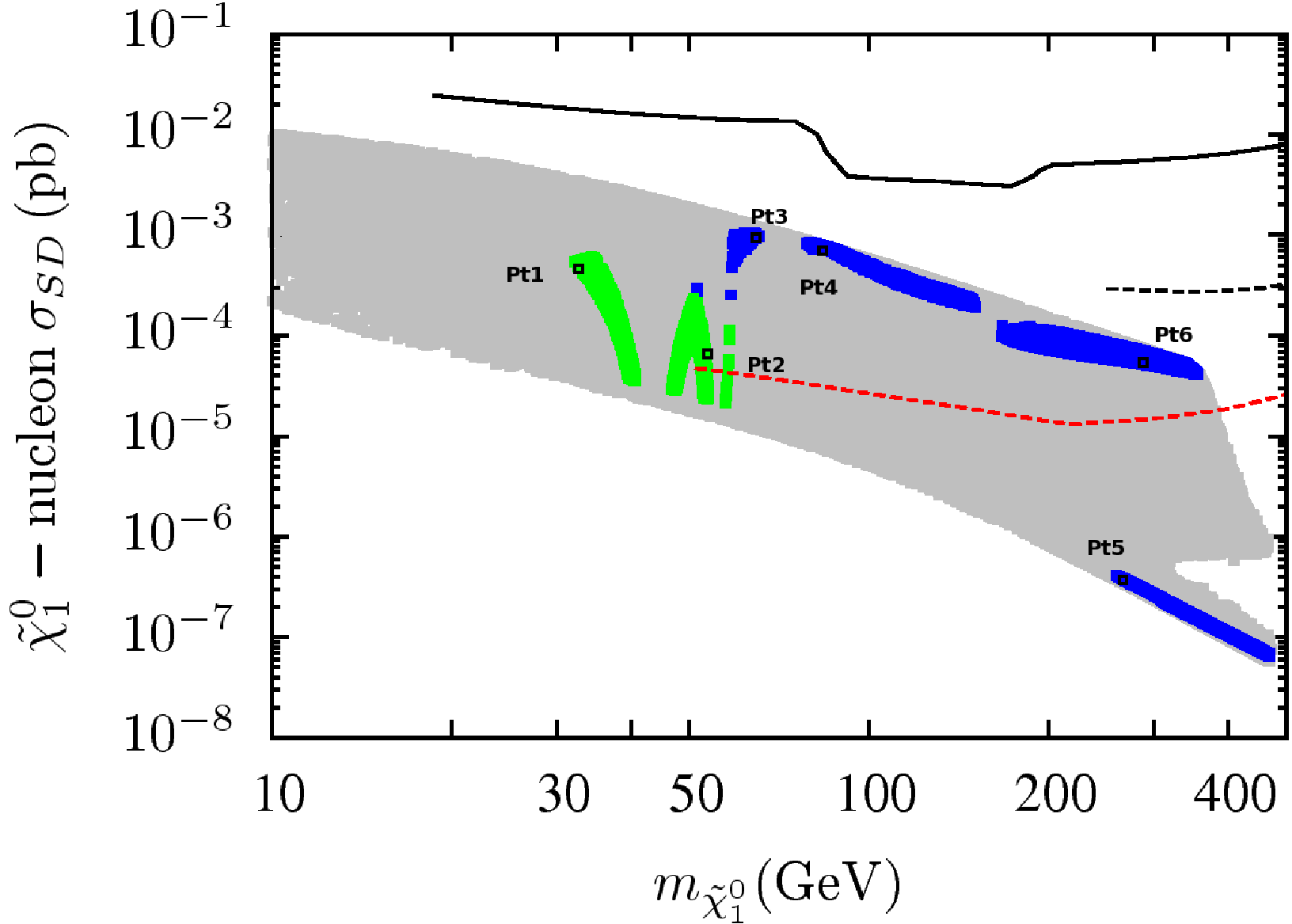}
}
\caption{
Spin-dependent elastic scattering cross section 
 versus neutralino dark matter mass. 
Color coding is the same as in Figure~\ref{m0M12A0TB30}. 
Current bounds from Super-Kamiokande (black line) 
 and IceCube (dotted black line), 
 and future reach of 
 IceCuce DeepCore (dotted red line) are  shown. 
Approximate locations of benchmark points 
listed in Tables~\ref{table5} and \ref{table6} are also shown. 
}
\label{SDA0TB50}
\end{figure}

Plots analogous to Figures~\ref{m0M12A0TB10}-\ref{SDA0TB10} 
 for varying values of $\tan \beta$ are shown 
 in Figures~\ref{m0M12A0TB30}-\ref{SDA0TB50}  
 and the particle mass spectra of the corresponding benchmark points
 are listed in Tables~\ref{table3}-\ref{table6}. 
 The figures for $ \mu < 0$ show green points 
 which belong to the subset of blue points and satisfy 
 all constraints including $\Delta(g-2)_\mu/2$, 
 where the deviation of the muon anomalous magnetic dipole moment 
 from the SM prediction can be explained by sparticle loop
 contributions. 
Since  the sparticle contributions to $\Delta(g-2)_\mu/2$ 
 are dominated by loop diagrams with chargino 
 and are proportional to $ \mu M_{2} \tan\beta/{\tilde m}_{SUSY}^4$,  
 where ${{\tilde m}_{SUSY}}$ is the sparticle mass in the loop, 
 the relative sign of $\mu$ and $M_2$ should be positive 
 ($\mu < 0$ in our convention) 
 in order to obtain $\Delta(g-2)_\mu/2 > 0$. 
 If the sparticles running in the loop diagrams are heavy, 
 a relatively large $\tan \beta$ is necessary 
 to satisfy the constraint from $\Delta(g-2)_\mu/2$. 
There is no green point in Figure~\ref{m0M12A0TB10} 
 because $\tan \beta =10$ is too small to be 
 compatible with the $\Delta(g-2)_\mu/2$ constraint. 
We also list benchmark points 
 (5 and 6 in Tables~\ref{table3}-\ref{table6}) 
 with relatively large neutralino mass, 
 some of which can/cannot be tested 
 by the future IceCube DeepCore experiment.

\begin{figure}
\centering
\subfigure{
\includegraphics[width=8cm]{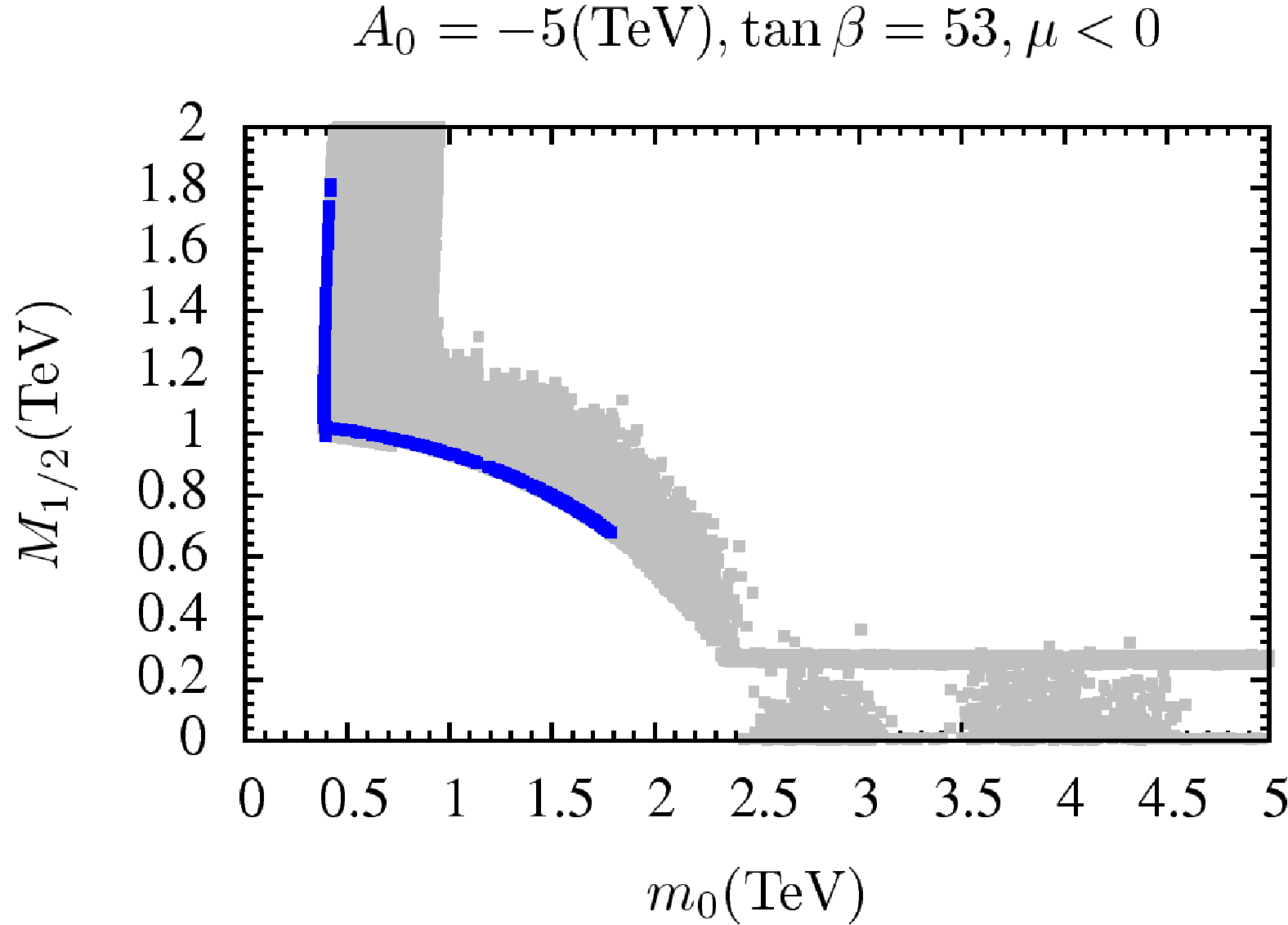}
} \subfigure{
\includegraphics[width=8cm]{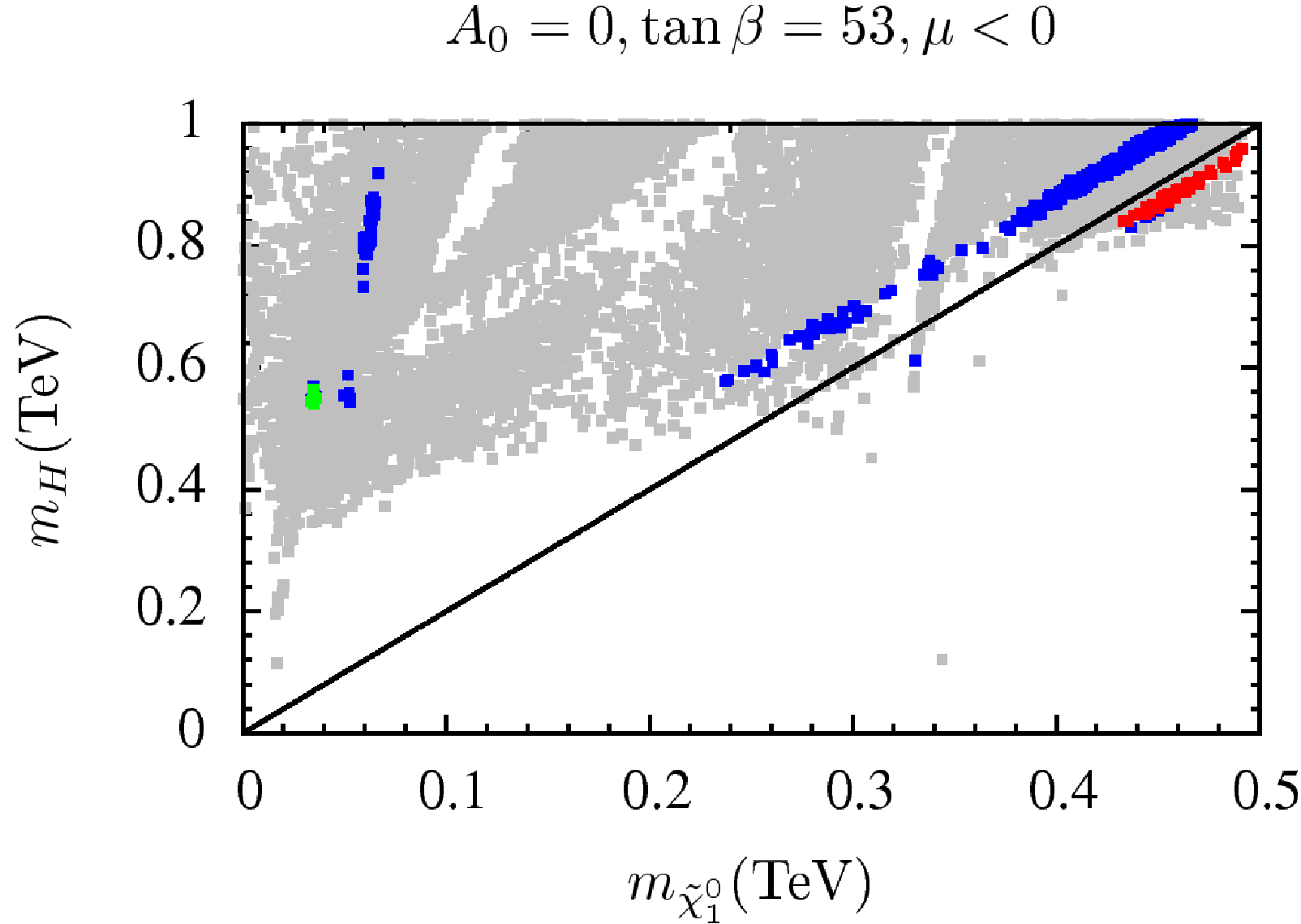}
} 
\caption{
Plots in the ($M_{1/2}, m_0$) and ($m_H, m_{\tilde{\chi}_1^{0}}$) planes.
Gray points are consistent with REWSB and $\tilde{\chi}^0_{1}$  
LSP. Blue points satisfy the WMAP bounds  on $\tilde{\chi}^0_1$ dark
matter abundance, particle mass bounds, 
 constraints from $BR(B_s\rightarrow \mu^+ \mu^-)$, 
 $BR(b\rightarrow s \gamma)$ and $BR(B_u\to\tau\nu_\tau)$. 
Green points belong to the subset of blue points 
 that satisfies all constraints including 
 $\Delta(g-2)_\mu/2$. 
Red points satisfy criterion of 
 $\vert 2m_{\tilde{\chi}_1^{0}}$- $m_H\vert \leq 30$.
}
\label{A5TB53}
\end{figure}

For $|A_0| \leq {\cal O}$(1 TeV), the results remain 
 almost the same as those with $A_0=0$, and 
 we therefore do not show all plots. 
The results for special cases with a large $A_0$ or 
 a large $\tan \beta$ are depicted in Figure~\ref{A5TB53}. 
The upper panel shows the results in the ($M_{1/2}, m_0$) plane 
 for $A_0=-5$ TeV, $\tan \beta =10$ and $\mu < 0$. 
An almost identical plot is also obtained for the opposite sign 
 of $\mu > 0$. 
There are two interesting regions. 
One is the usual stau co-annihilation region 
 for $M_{1/2} \gtrsim 1$ TeV, 
 the other one is the region for $M_{1/2} \lesssim 1$ TeV 
 where the mass difference between the neutralino LSP and 
 lighter stop is small, and the desired relic  abundance 
 of  neutralino dark matter is achieved through 
 stop-neutralino co-annihilations. 
This stop-neutralino co-annihilation region only appears for 
 a big negative $A_0$. 
The lower panel corresponds to the results for 
 a relatively large $\tan \beta( =53)$ in the ($m_H, m_{\tilde{\chi}_1^{0}}$) plane. 
The allowed region appears for $2 m_{\tilde{\chi}_1^{0}} \simeq m_H$ 
 by the enhancement of neutralino annihilation cross sections 
 via heavy Higgs boson exchange processes in the $s$-channel. 
For the chosen benchmark points, the mass spectra are listed 
 in Table~\ref{table7}.

Finally, we choose several benchmark points from the allowed 
 regions for different values of $A_0$, $\tan \beta$ 
 and ${\rm sign}(\mu)$ and compare the mass spectra in our  NUGM model  
with those from the CMSSM. 
In Table~\ref{table8}, we present four benchmark points 
 from the stau-neutralino co-annihilation region 
to compare the mass spectra. 
For the same values of $A_0$, $\tan \beta$ and ${\rm sign}(\mu)$,
 we tune $m_0$ and $M_{1/2}$ in the CMSSM in order to obtain 
 the same masses for neutralino LSP and the lighter stau as those found in NUGM model. 
The resultant masses for the other particles are all 
 larger than those in the CMSSM. 
In Table~\ref{table9}, we tune $m_0$ and $M_{1/2}$ in the CMSSM 
 so as to give the same masses for gluino and right-handed 
 down squarks as in the NUGM model. 
We can see, in this case, large mass differences 
 in the neutralino and chargino mass spectra, while 
the sfermion mass spectra are similar. 
In particular, the mass of the neutralino LSP is 
 relatively small as a result of the non-universal boundary 
 conditions for the gaugino masses. 
The results in these Tables show that the 
 mass spectra in our NUGM model are quite distinct from those in the CMSSM.

\section{Conclusions\label{conclusions}}

We have investigated the low scale particle spectroscopy arising from supersymmetric SO(10) and SU(5) models 
 with non-universal gaugino masses. 
This non-universality generally arises from the $F$-term 
 of some non-singlet hidden sector field in the gauge kinetic function 
 in the gravity mediated supersymmetry breaking. 
Depending on the embedding of the MSSM gauge group, 
 the ratio of the MSSM gaugino masses is determined 
 by group theory. 
Among several possibilities, we have considered 
 an $F$-term from a single {\bf 54}-plet hidden sector field. 
This is a unique possibility  if we require 
 the SO(10) gauge coupling to stay within the perturbative regime 
 up to the Planck scale. 
With the group theoretically fixed ratios of the  
 gaugino masses, we set the fundamental parameters 
 of the model in the same way as the CMSSM, namely, 
 $m_0$, $M_{1/2}$, $A_0$, $\tan \beta$ and ${\rm sign(\mu)}$ 
 with the identification $M_3=M_{1/2}$. 
Taking a variety of phenomenological constraints into account, 
 we have identified the allowed parameter regions. 
We have also calculated the spin-independent and dependent 
 cross sections of neutralino elastic scattering 
 off a nucleon and compared the results with 
 the reach of future direct and indirect detection experiments. 
We have identified the benchmark points of the model 
 which are consistent with the phenomenological  constraints 
 and which predict neutralino elastic scattering cross sections 
 accessible at the future dark matter detection experiments. 
The particle mass spectra of the benchmark points  
 can be well-distinguished at the LHC from the mass spectra in the CMSSM.

\begin{spacing}{2.5}

{\noindent}{\bf Note Added}: $t$-$b$-$\tau$ Yukawa unification in this class of models with $\mu < 0$ has recently been investigated in \cite{GSU}.

\end{spacing}


\acknowledgments

We thank Ilia Gogoladze and Rizwan Khalid for useful comments and discussion. 
This work is supported in part by the DOE Grants, 
No. DE-FG02-10ER41714 (N.O.) and No. DE-FG02-91ER40626 (Q.S. and S.R.) 
and by Bartol Research Institute (S.R.). 
N.O. would like to thank the Particle Theory Group of 
 the University of Delaware for hospitality during his visit.

\newpage

\begin{figure}
\centering
\subfiguretopcaptrue
\subfigure[\hspace {1mm} $A_0=0,\tan\beta=10,\mu>0$]{
\includegraphics[width=8cm]{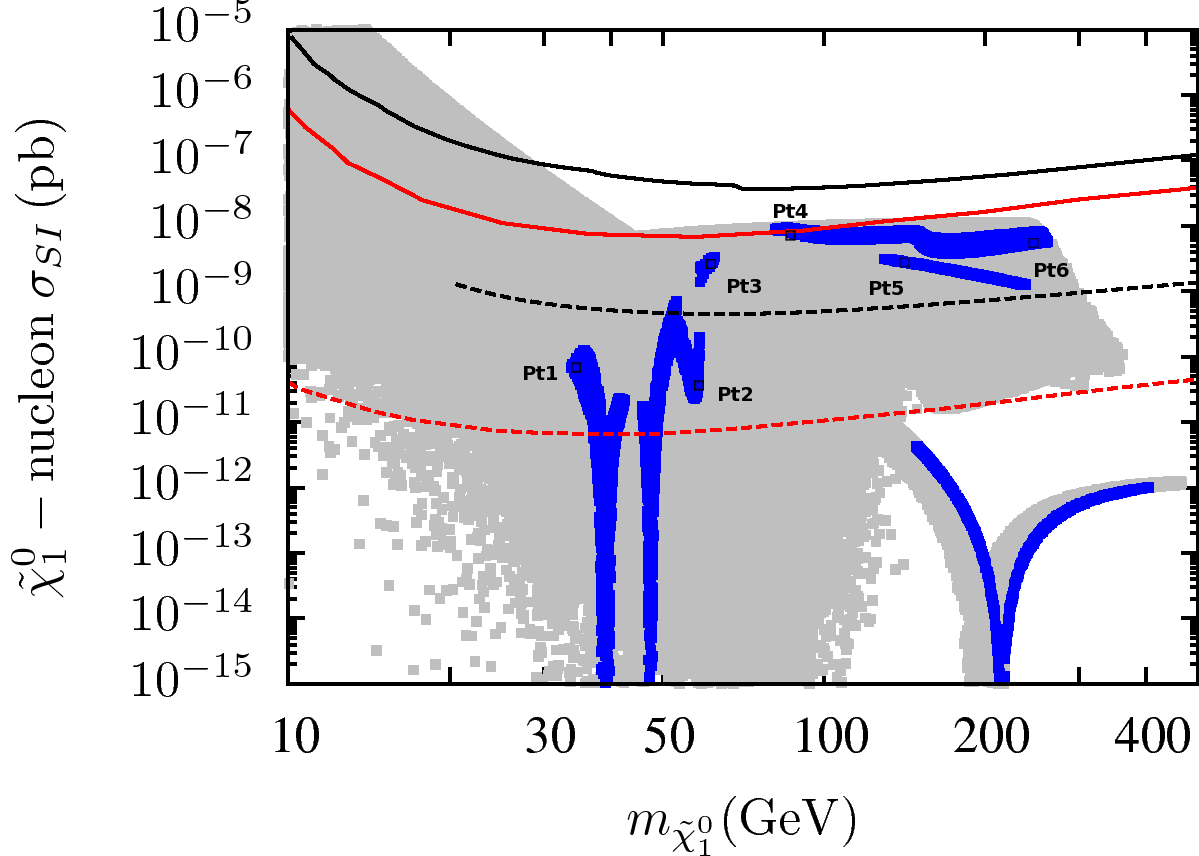}
} \subfigure[\hspace {1mm} $A_0=0,\tan\beta=10,\mu<0$]{
\includegraphics[width=8cm]{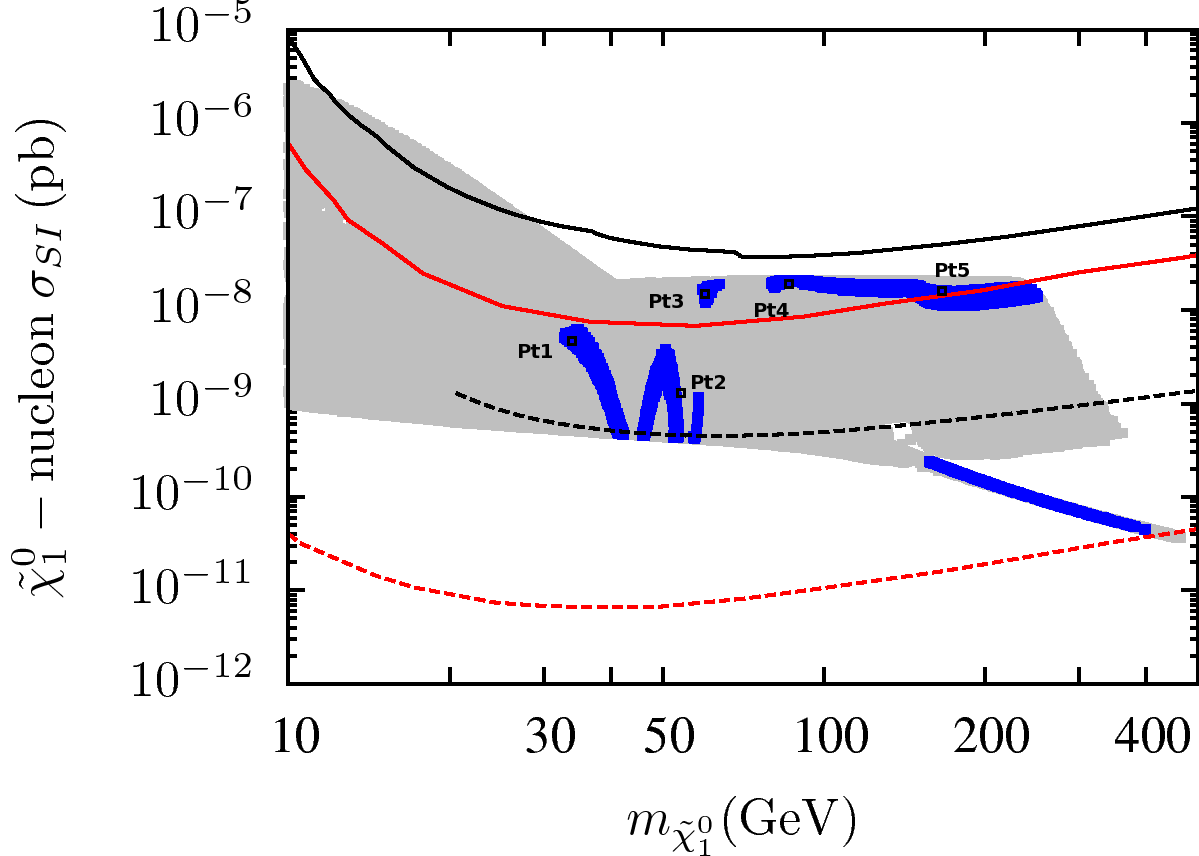}
}
\caption{ 
Spin-independent elastic scattering cross section 
 of neutralino dark matter 
 in the ($\sigma_{\rm SI}, m_{\tilde{\chi}_1^{0}}$) plane. 
Color coding is the same as in Figure~\ref{m0M12A0TB10}. 
The current upper bounds 
 from CDMS-II (XENON100) are depicted as black (red) solid lines. 
 Future reach of the SuperCDMS(SNOLAB) (dotted black line) 
 and XENON1T (dotted red line) are shown.  
Approximate locations of benchmark points 
 listed in Tables~\ref{table1} and \ref{table2} are also shown, and which are
testable in the ongoing experiments. 
}
\label{SIA0TB10}
\end{figure}

\begin{figure}
\centering
\subfiguretopcaptrue
\subfigure[\hspace {1mm} $A_0=0,\tan\beta=10,\mu>0$]{
\includegraphics[width=8cm]{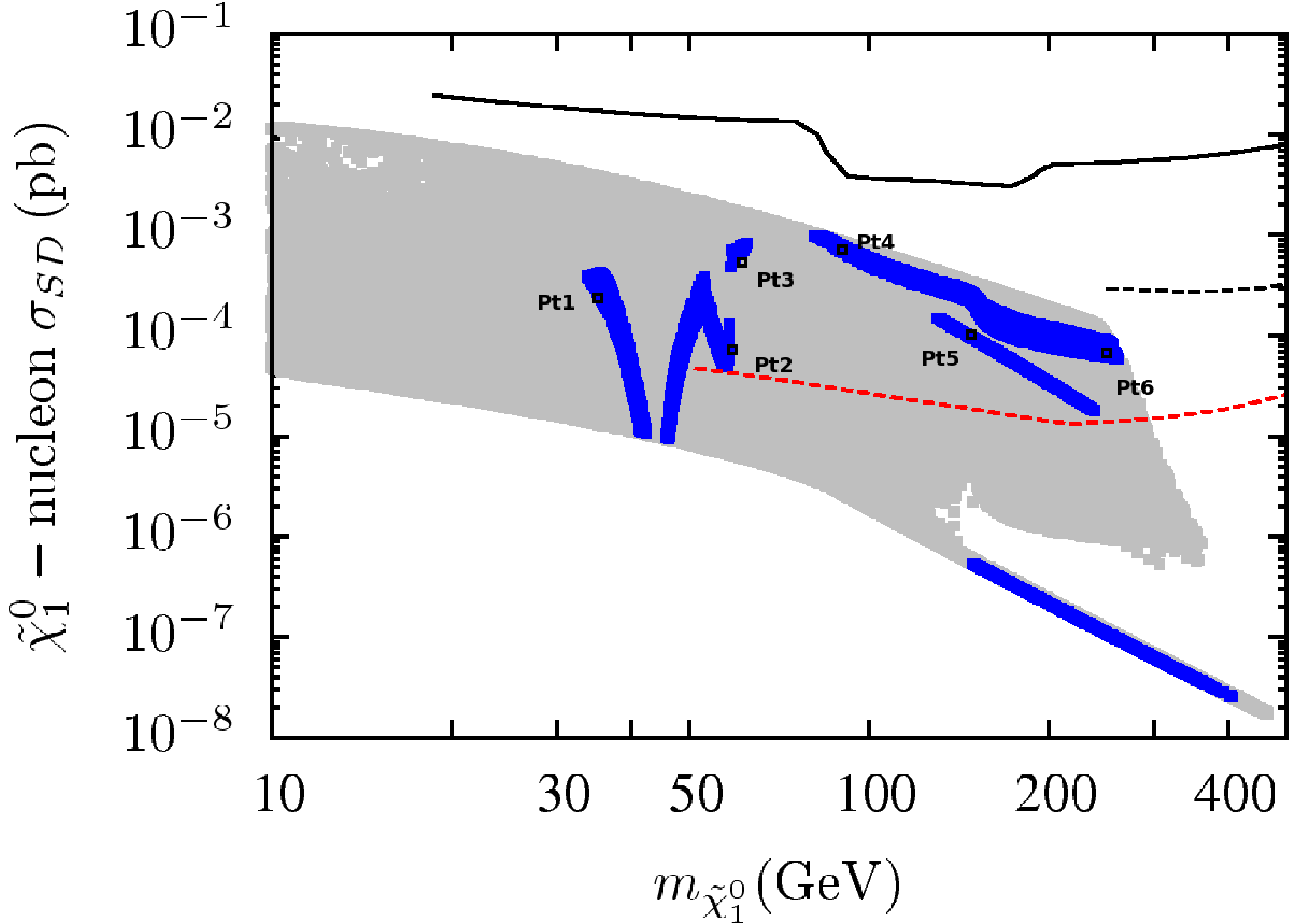}
} \subfigure[\hspace {1mm} $A_0=0,\tan\beta=10,\mu<0$]{
\includegraphics[width=8cm]{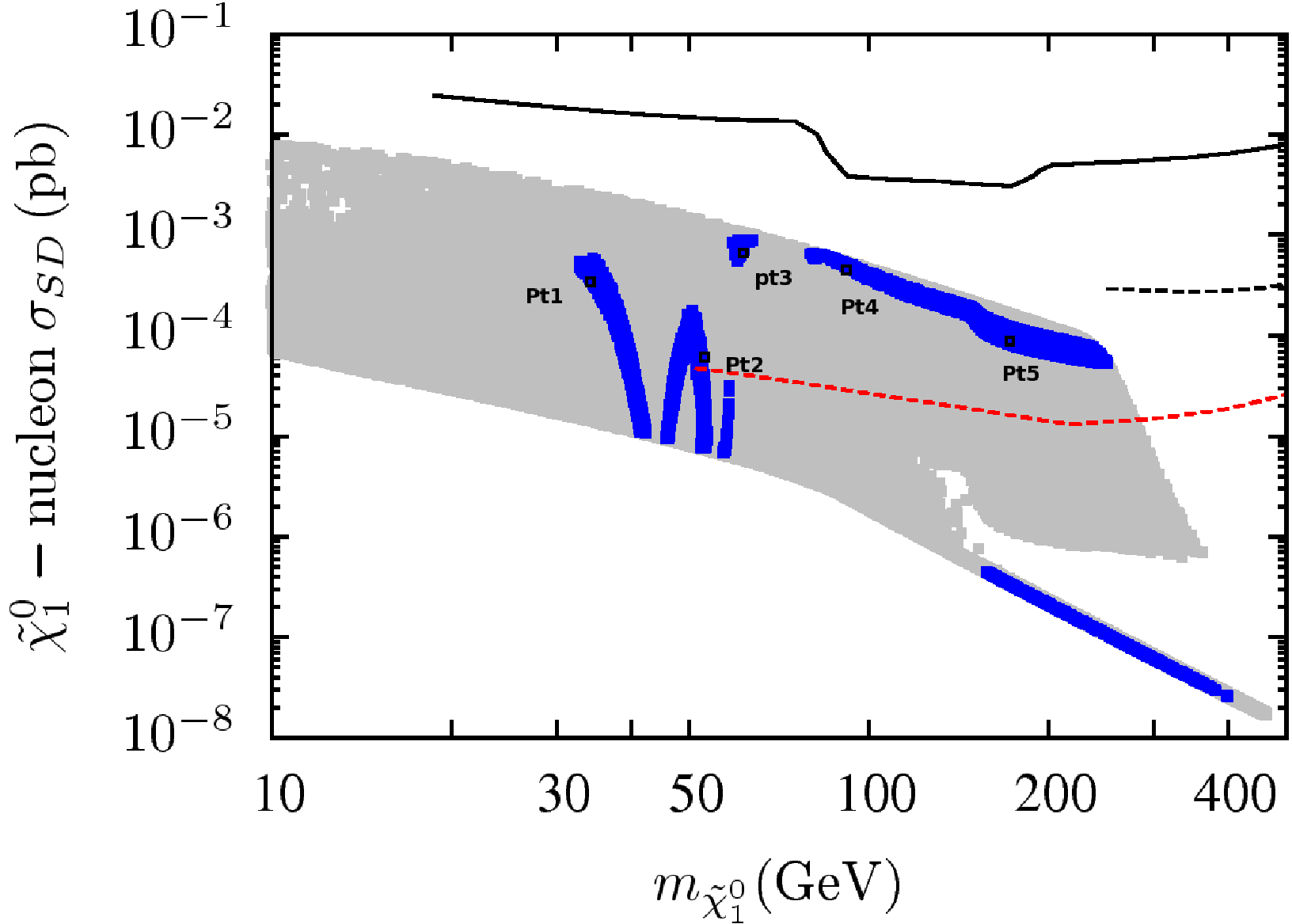}
}

\caption{
Spin-dependent elastic scattering cross section
of neutralino dark matter 
in the ($\sigma_{\rm SD}, m_{\tilde{\chi}_1^{0}}$) plane. 
Color coding is the same as in Figure~\ref{m0M12A0TB10}.
Current bounds from Super-Kamiokande (black line),
IceCube (dotted black line), and future reach of IceCube DeepCore experiment 
(dotted red line) are shown. Approximate locations of benchmark points presented in Tables~\ref{table1} and \ref{table2} are also shown.
}

\label{SDA0TB10}
\end{figure}

\begin{table}
\centering
\begin{tabular}{lcccccc}
\hline
\hline
               &Pt1  & Pt2 & Pt3    & Pt4     &  Pt5 & Pt6   \\
\hline
$m_{0}$        & 2948 & 3134 & 3404  & 3780 &  4384  & 4982   \\
$M_{1/2} $     & 154 &  250  & 292   & 447  & 778   & 1166   \\
$A_0 $         & 0 &  0      & 0     &   0 & 0     &  0     \\
$sign(\mu) $   & + &  +      & +    &   + & +     &   +   \\
$\tan\beta$    &10  & 10  &  10    &   10 &  10   &   10   \\

\hline
$m_h$          &116  & 117 & 118   & 118  & 119 & 119      \\
$m_H$          &2926  & 3120 & 3387  & 3773 & 4418 & 5078     \\
$m_A$          &2907  & 3099 & 3365  & 3749 & 4389 & 5045   \\
$m_{H^{\pm}}$  &2927  & 3121 & 3388  & 3774 & 4419  & 5079    \\

\hline
$m_{\tilde{\chi}^0_{1,2}}$
               &34,108  & 56, 176   & 59,119   & 81, 121  & 170, 234 & 262, 323   \\
$m_{\tilde{\chi}^0_{3,4}}$
               &138,239  & 197, 355  & 127,404  & 132, 603 & 238, 1032 & 330,1537  \\

$m_{\tilde{\chi}^{\pm}_{1,2}}$
               &104,233  & 168, 345  & 106,391  & 104, 585  & 213, 1006 & 297, 1502  \\
$m_{\tilde{g}}$ & 490 &    733     &  838     & 1207 & 1951  & 2784    \\

\hline $m_{ \tilde{u}_{L,R}}$
               &2936,2946  &3150, 3154  & 3429,3432 & 3863,3855 &4637, 4596  & 5486,5395     \\
$m_{\tilde{t}_{1,2}}$
               &1687,2402  & 1824, 2587  & 1994,2820 & 2280,3196 &2819, 3884  & 3433,4657   \\                                               
\hline $m_{ \tilde{d}_{L,R}}$
               &2938,2948  & 3151, 3156  & 3430,3434 & 3864,3857 &4638, 4598  & 5487,5397    \\
$m_{\tilde{b}_{1,2}}$
               &2395,2921  & 2577, 3127  & 2809,3403 & 3184,3822 &3871, 4556 & 4641,5348  \\
\hline
$m_{\tilde{\nu}_{1}}$
               &2946  &    3137        &  3410       &  3798    & 4441     & 5099    \\
$m_{\tilde{\nu}_{3}}$
               &2933  &    3124        &  3395       &  3782  &   4423   & 5078   \\
\hline
$m_{ \tilde{e}_{L,R}}$
              &2946,2945  & 3137, 3131     & 3409,3401     & 3798,3777 &4440, 4382  & 5097,4982   \\
$m_{\tilde{\tau}_{1,2}}$
              &2919,2933  & 3104, 3123       & 3372,3394   & 3745,3781  &4345, 4421   & 4940,5076  \\
\hline
$\sigma_{SI}({\rm pb})$
              & $6.8\times 10^{-11}$  & $6.2 \times 10^{-11}$ & $2.4 \times 10^{-9}$ & $8.9 \times 10^{-9}$ & $4.8 \times 10^{-9}$ & $5.7 \times 10^{-9}$ \\

$\sigma_{SD}({\rm pb})$
              & $3.8 \times 10^{-4}$  & $8.2\times 10^{-5}$ & $7.1\times 10^{-4}$ & $9.9\times 10^{-4}$ & $1.2\times 10^{-4}$ & $6.0\times 10^{-5}$ \\

$\Omega_{CDM}h^2$
            &0.13   & 0.1  &  0.10  &  0.12 & 0.1 & 0.13 \\
\hline
\hline
\end{tabular}
\caption{
Mass spectra for the benchmark points with $\tan\beta$=10, $\mu> 0$. All of these points satisfy the various constraints
mentioned in section~\ref{constraintsSection}, except  $\Delta(g-2)_\mu/2$. Neutralino
LSP in all cases has sizeable higgsino component. The points lie between current and future direct and indirect dark matter search limits
shown in Figures~\ref{SIA0TB10} and \ref{SDA0TB10}. Pt1 represents the lightest neutralino and a light gluino. Pts.2 and 3 also have relatively light
gluinos.
\label{table1}}
\end{table}
\begin{table}
\centering
\begin{tabular}{lccccc}
\hline
\hline
                 & Pt1 & Pt2    & Pt3     &  Pt4 & Pt5   \\
\hline
$m_{0}$          & 3071 & 3061  & 3573  & 3984   & 4490   \\
$M_{1/2} $       & 184   & 249   & 358 &  536   & 855   \\
$A_0 $           &  0    & 0     &   0 & 0    &  0     \\
$sign(\mu) $     &  -    & -     &   - & -    &   -   \\
$\tan\beta$      & 10  &  10    &   10 &  10   &   10   \\

\hline
$m_h$            & 117 & 117   & 118  & 118 & 119      \\
$m_H$            & 3050 & 3048  &3560   & 3986  & 4535     \\
$m_A$            & 3030 & 3028  & 3536 & 3960  &4505    \\
$m_{H^{\pm}}$    & 3051  & 3049  & 3560 & 3986  & 4536    \\

\hline
$m_{\tilde{\chi}^0_{1,2}}$
                 &33,110   &52, 184   &62 ,122   &98, 147 &182 ,252   \\
$m_{\tilde{\chi}^0_{3,4}}$
                 &134 ,278  &209, 362  &128 ,490 &163, 720 &262 ,1133  \\

$m_{\tilde{\chi}^{\pm}_{1,2}}$
                 & 106 ,271  &185, 355  &112 ,478  &143, 705 &255 ,1113  \\
$m_{\tilde{g}}$  & 567  & 730  & 998  & 1424  & 2119    \\

\hline $m_{ \tilde{u}_{L,R}}$
                 &3066 ,3074  & 3078, 3082 & 3621, 3620 &4107, 4091  &4791 ,4740  \\
$m_{\tilde{t}_{1,2}}$
                 &1765, 2512  & 1783, 2528 &2119, 2985 &2443, 3408 &2932 ,4028  \\                                                                                   
\hline $m_{ \tilde{d}_{L,R}}$
                 & 3067 ,3076  & 3080, 3084 &3620, 3622  &4108, 4093 &4792 ,4742    \\
$m_{\tilde{b}_{1,2}}$
                 & 2502, 3048  & 2518, 3056 &2974, 3588 &3396, 4056 &4011, 4698  \\
\hline
$m_{\tilde{\nu}_{1}}$
                 &  3070          &  3065       & 3584     & 4011     & 4558    \\
$m_{\tilde{\nu}_{3}}$
                 &  3057          &  3052       & 3568   &  3994   & 4540   \\
\hline
$m_{ \tilde{e}_{L,R}}$
                & 3070 ,3068     & 3065, 3058     &3583  ,3570 & 4011, 3981 &4557  , 4488  \\
$m_{\tilde{\tau}_{1,2}}$
                & 3041 ,3056       &3017, 3051   &3539  ,3567  & 3947, 3993   &4450  ,4538  \\
\hline
$\sigma_{SI}({\rm pb})$
                & $5.1 \times 10^{-9}$ & $1.7 \times 10^{-9}$ & $1.5 \times 10^{-8}$ & $1.9 \times 10^{-8}$ & $1.2 \times 10^{-8}$ \\

$\sigma_{SD}({\rm pb})$
                & $4.8\times 10^{-4}$ & $6.5\times 10^{-5}$ & $7.1\times 10^{-4}$ & $4.1\times 10^{-4}$ & $7.9\times 10^{-5}$ \\

$\Omega_{CDM}h^2$
               & 0.13  &  0.11  &  0.13 & 0.11 & 0.12 \\
\hline
\hline
\end{tabular}
\caption{ 
Mass spectra for the benchmark points for $\tan\beta$=10, $\mu < 0$. All of these points satisfy the various constraints
mentioned in section~\ref{constraintsSection}, except  $\Delta(g-2)_\mu/2$. Neutralino
LSP in all cases has sizeable higgsino component. The points lie between present and future direct and indirect dark matter search limits
shown in Figures~\ref{SIA0TB10} and \ref{SDA0TB10}. Pt1 represents lightest neutralino, while Pts.1,2,3 have relatively light
gluinos.
\label{table2}}
\end{table}

\clearpage

\begin{table}
\centering
\begin{tabular}{lcccccc}
\hline
\hline
                &Pt1 & Pt2 & Pt3    & Pt4  & Pt5  &  Pt6    \\
\hline
$m_{0}$         &2038 & 2158  &2616   & 2980 & 201   & 4382      \\
$M_{1/2} $      &166  & 245   & 341   & 488  & 650   & 1280       \\
$A_0 $          &0 &  0    & 0     &   0     & 0     & 0          \\
$sign(\mu) $    &+  &  +   & +   &   +    & +    & +       \\
$\tan\beta$     &30 & 30  &  30    &    30   &  30   &  30      \\

\hline
$m_h$          &115  & 116  & 117     & 118   & 115  & 120      \\
$m_H$          &1734  & 1852  & 2247   & 2578 & 800  & 3939      \\
$m_A$          &1723  & 1840  & 2232   & 2561 & 795  & 3914      \\
$m_{H^{\pm}}$  & 1737 & 1853  & 2248   & 2579 & 804  & 3940      \\

\hline
$m_{\tilde{\chi}^0_{1,2}}$
               &34, 110 &52, 169  & 64, 123  & 91, 137 & 144, 581 & 283, 337    \\
$m_{\tilde{\chi}^0_{3,4}}$
               &138, 250  &189, 345   & 124, 463 &148, 651 & 596, 831 & 349, 1677   \\

$m_{\tilde{\chi}^{\pm}_{1,2}}$
               &104, 245  &160, 337   & 103, 450  &119, 633 & 565, 818  & 311, 1641  \\
$m_{\tilde{g}}$ & 503 &   696       &  934      & 1276 & 1486  & 2990      \\

\hline $m_{ \tilde{u}_{L,R}}$
               & 2049, 2052 &2202, 2199   & 2694, 2685  &3132, 3109 & 1444, 1301  &5077, 4949      \\
$m_{\tilde{t}_{1,2}}$
               &1184, 1604  &1289, 1738    & 1595, 2137 &1888, 2510 & 1068, 1355  &3264, 4235      \\
\hline $m_{ \tilde{d}_{L,R}}$
               &2051, 2053  &2204, 2201   & 2695, 2686  &3133, 3110 & 1446, 1300  &5077, 4950       \\
$m_{\tilde{b}_{1,2}}$
               &1593, 1903  &1727, 2044    & 2125, 2499  &2497, 2901 & 1258, 1341  &4217, 4667     \\
\hline
$m_{\tilde{\nu}_{1}}$
               &2040  &     2166       &  2632       &   3012   & 666  &  4544      \\
$m_{\tilde{\nu}_{3}}$
               &1960  &     2082       & 2531        &   2898   & 650  &  4381       \\
\hline
$m_{ \tilde{e}_{L,R}}$
              &2041, 2036  &2167, 2157     & 2632, 2614    &3012, 2978 &673, 233  &4544, 4384      \\
$m_{\tilde{\tau}_{1,2}}$
              &1874, 1960  &1985, 2083      & 2408, 2531   &2744, 2898 &152, 659  &4038, 4379     \\
\hline
$\sigma_{SI}({\rm pb})$
               &$1.6 \times 10^{-10}$ & $1.5\times 10^{-10}$ & $5.2\times 10^{-9}$ & $9.5\times 10^{-9}$ & $6.0\times 10^{-11}$ & $8.9\times 10^{-9}$  \\

$\sigma_{SD}({\rm pb})$
              & $4.2 \times 10^{-4}$ & $1.0\times 10^{-4}$ & $8.4\times 10^{-3}$ & $6.5\times 10^{-4}$ & $7.0\times 10^{-7}$ & $5.9\times 10^{-5}$  \\

$\Omega_{CDM}h^2$
               & 0.13 & 0.11        & 0.13   &  0.1  & 0.1 & 0.1 \\
\hline
\hline
\end{tabular}
\caption{
Mass spectra for the benchmark points for $\tan\beta$=30, $\mu> 0$. All of these points satisfy the various constraints
mentioned in section~\ref{constraintsSection}, except  $\Delta(g-2)_\mu/2$. For all points, the neutralino
LSP has sizeable higgsino component. The points lie between present and future direct and indirect dark matter search limits
shown in Figures~\ref{SIA0TB30} and \ref{SDA0TB30}. Pt1 represents lightest neutralino scenario and 
Pt5 shows stau-coannihilation scenario. 
\label{table3}}
\end{table}

\begin{table}
\centering
\begin{tabular}{lcccccc}
\hline
\hline
                 & Pt1 & Pt2    & Pt3     &  Pt4 & Pt5 & Pt6  \\
\hline
$m_{0}$          &1939  & 2510  &2790  & 3070   & 219 & 4490 \\
$M_{1/2} $       & 183   & 311   & 402 &  525  & 616  &  1340  \\
$A_0 $           &  0    & 0     &   0 & 0    &  0    &   0  \\
$sign(\mu) $     &  -   & -   &   - & -    &   - &  -  \\
$\tan\beta$      & 30  &  30    &   30 &  30   &   30 &  30  \\

\hline
$m_h$            &115  & 117   & 118  & 119    & 115   & 121  \\
$m_H$            &1640  & 2139  &2390  & 2650  & 695   & 400 \\
$m_A$            &1629  & 2125  &2374 &  2630  & 690   & 3970 \\
$m_{H^{\pm}}$    &1642  & 2141  & 2391 & 2650  & 700  &  4000\\

\hline
$m_{\tilde{\chi}^0_{1,2}}$
                 &37,142   &59 ,130   & 66,112  &96, 142 &136 ,553 &290, 337   \\
$m_{\tilde{\chi}^0_{3,4}}$
                 &177 ,279  &135,425  &126 ,541 &157, 700 &569,791 &355, 1750 \\

$m_{\tilde{\chi}^{\pm}_{1,2}}$
                 &142 ,276  & 123,417  &104 ,530  &138, 686 &559, 784 &344, 1720  \\
$m_{\tilde{g}}$  &  542       &  861     & 1078 & 1360  & 1420    &3100 \\

\hline $m_{ \tilde{u}_{L,R}}$
                 & 1960,1961  &2576 ,2568 & 2896 ,2881 &3240, 3200 &1380, 1240 &5200, 5090 \\
$m_{\tilde{t}_{1,2}}$
                 &1137 ,1534  &1519 ,2035 & 1728 ,2303 &1960, 2600 &1020, 1280 &3370, 4350  \\
\hline $m_{ \tilde{d}_{L,R}}$
                 &1962 ,1963  &2577 ,2570 &2897 ,2884 &3240, 3220 &1380, 1240 &5230, 5090   \\
$m_{\tilde{b}_{1,2}}$
                 &1524 ,1810  &2023 ,2382 &2291 ,2679 &2590, 2990 &1170, 1270 &4330, 4780 \\
\hline
$m_{\tilde{\nu}_{1}}$
                 &  1942          & 2523        &  2812    & 3110     & 641  & 4660 \\
$m_{\tilde{\nu}_{3}}$ & 1865      & 2425        & 2704     & 2990      & 622 & 4490  \\
\hline
$m_{ \tilde{e}_{L,R}}$
                &1944 , 1938    &2523 ,2508     &2812 ,2788 &3110, 3070 & 648, 247  &4660, 4490 \\
$m_{\tilde{\tau}_{1,2}}$
                &1779 ,1866       &2307 ,2425   &2566 ,2703  &2820, 2990   &145,633 & 4130, 4490\\
\hline
$\sigma_{SI}({\rm pb})$
                & $1.9 \times 10^{-9}$ & $8.6 \times 10^{-9}$ & $1.8 \times 10^{-8}$ & $1.7 \times 10^{-8}$ & $7.9 \times 10^{-10}$ &$1.5 \times 10^{-8}$ \\

$\sigma_{SD}({\rm pb})$
                & $1.4\times 10^{-4}$ & $5.3\times 10^{-4}$ & $1.0\times 10^{-3}$ & $5.3\times 10^{-4}$ & $8.4\times 10^{-7}$ & $6.0\times 10^{-5}$ \\

$\Omega_{CDM}h^2$
               & 0.13  &  0.12  &  0.13 & 0.1 & 0.13 & 0.1 \\
\hline
\hline
\end{tabular}
\caption{
Mass spectra for the benchmark points for $\tan\beta$=30, $\mu< 0$. All of these points satisfy the various constraints
mentioned in section~\ref{constraintsSection}. For all points, the neutralino
LSP has sizeable higgsino component. The points lie between present and future direct and indirect dark matter
search limits shown in Figures~\ref{SIA0TB30} and \ref{SDA0TB30}. Pt1 shows the lightest neutralino and also a light gluino. 
Pt5 shows a stau-coannihilation scenario. 
\label{table4}}
\end{table}

\clearpage


\begin{table}
\centering
\begin{tabular}{lcccccc}
\hline
\hline
                 & Pt1 & Pt2    & Pt3     &  Pt4 & Pt5 & Pt6 \\
\hline
$m_{0}$          & 2241 & 1851  & 3175  & 3529  & 392  & 4765 \\
$M_{1/2} $       & 167   & 247  & 489 & 783     & 735   & 1440  \\
$A_0 $           &  0    & 0     &   0 & 0      &  0    &  0  \\
$sign(\mu) $     &  +   & +   &   + & +     &   +   & +    \\
$\tan\beta$      & 50  &  50    &   50 &  50    &   50   & 50   \\

\hline
$m_h$            & 116 & 115   & 119  & 120  & 116   & 121   \\
$m_H$            & 1046 & 967  & 1609 & 1944 & 750   & 2802   \\
$m_A$            & 1039 & 961  & 1598 & 1932 & 745   & 2785 \\
$m_{H^{\pm}}$    & 1050 & 972  & 1613 & 1947 & 755   & 2803   \\

\hline
$m_{\tilde{\chi}^0_{1,2}}$
                 &33, 111   &54, 225   &90, 136  &175, 287 &165, 642  & 320, 371\\
$m_{\tilde{\chi}^0_{3,4}}$
                 &138, 252  &257, 357  &148, 654 &287 ,1032 &655, 938 &386, 1886 \\

$m_{\tilde{\chi}^{\pm}_{1,2}}$
                 &104, 247  &218, 350  &119, 636  &264, 1007 &623, 923 &344, 1845  \\
$m_{\tilde{g}}$  &  509       &  693     & 1284  & 1928  & 1669  & 3326 \\

\hline $m_{ \tilde{u}_{L,R}}$
                 &2249, 2253  &1913, 1906 &3317, 3296 &3868, 3808 &1649, 1490 &5564, 5416    \\
$m_{\tilde{t}_{1,2}}$
                 &1300, 1578  &1130, 1387 &1993, 2423 &2407, 2934 &1210, 1508 &3593, 4386  \\
\hline $m_{ \tilde{d}_{L,R}}$
                 &2251, 2255  &1914, 1907 &3318, 3298 &3868, 3809 &1651, 1489 &5565, 5417   \\
$m_{\tilde{b}_{1,2}}$
                 &1564, 1777  &1373, 1545 &2407, 2672 &2917, 3170 &1375, 1496 &4362, 4614 \\
\hline
$m_{\tilde{\nu}_{1}}$
                 &      2242      &  1862       &  3205    & 3604     & 817  & 4954 \\
$m_{\tilde{\nu}_{3}}$
                 &     1984       &  1654       & 2850   &  3217      & 770  & 4444 \\
\hline
$m_{ \tilde{e}_{L,R}}$
                &2243, 2239    & 1864, 1850     &3206, 3173 & 3604, 3528   & 823, 414 & 4953, 4768  \\
$m_{\tilde{\tau}_{1,2}}$
                &1683, 1984    &1397, 1654      &2400, 2859  &2677 , 3215  & 173, 779 &3621, 4440  \\
\hline
$\sigma_{SI}({\rm pb})$
                & $6.3 \times 10^{-10}$ & $4.5 \times 10^{-10}$ & $6.2 \times 10^{-9}$ & $1.4 \times 10^{-9}$ & $2.5 \times 10^{-10}$ & $8.9 \times 10^{-9}$\\

$\sigma_{SD}({\rm pb})$
                & $4.2\times 10^{-4}$ & $2.7\times 10^{-5}$ & $6.5\times 10^{-4}$ & $3.6\times 10^{-5}$ & $4.9\times 10^{-7}$ & $4.9\times 10^{-5}$ \\

$\Omega_{CDM}h^2$
               & 0.13  &  0.11  &  0.1 & 0.13 & 0.13 & 0.1\\ 
\hline
\hline
\end{tabular}
\caption{ 
 Benchmark points for $\tan\beta$=50, $\mu> 0$. All of these points satisfy the various constraints
mentioned in section~\ref{constraintsSection}, except  $\Delta(g-2)_\mu/2$. In all cases, the neutralino
LSP has sizeable higgsino component. The points lie between present and future direct and indirect searches of dark matter
shown in Figures~\ref{SIA0TB50} and \ref{SDA0TB50}. Pt1 represents lightest neutralino, and Pt5 shows stau-coannihilation scenario. Pts.1, and 2 show 
relatively light gluinos.
\label{table5}}
\end{table}

\begin{table}
\centering
\begin{tabular}{lcccccc}
\hline
\hline
                 & Pt1 & Pt2    & Pt3     &  Pt4 & Pt5 & Pt6   \\
\hline
$m_{0}$          & 2237 & 1944  &2893  & 3248   & 2856 & 4495 \\
$M_{1/2} $       & 174   & 246   & 368 & 509   & 1148  & 1259 \\
$A_0 $           &  0    & 0     &   0 & 0    &  0     & 0  \\
$sign(\mu) $     &  -   & -   &   - & -    &   -  & -  \\
$\tan\beta$      & 50  &  50    &   50 &  50   &   50  &  50  \\

\hline
$m_h$            & 116 & 115   & 118 & 119  &  119 &  121  \\
$m_H$            & 902 & 618  & 1276 & 1486 & 593  &  2200 \\
$m_A$            & 896 & 614  & 1267 & 1476 & 589  &  2185 \\
$m_{H^{\pm}}$    & 907 & 624  & 1279 & 1489 & 600  &  2202 \\

\hline
$m_{\tilde{\chi}^0_{1,2}}$
                 &34, 115   &53, 217   &64, 117   &84, 121 &265, 745 &276, 330   \\
$m_{\tilde{\chi}^0_{3,4}}$
                 &143, 264  & 251, 358 &124, 498  &145, 679 &749, 1487 &344, 1650  \\

$m_{\tilde{\chi}^{\pm}_{1,2}}$
                 &114, 259  &219, 355  &108, 487  &116, 666 &760, 1465 &336, 1625  \\
$m_{\tilde{g}}$  &   527      &  693     & 1001   &1330  & 2643 & 2949  \\

\hline $m_{ \tilde{u}_{L,R}}$
                 &2248, 2251  &2001, 1995 &2975, 2965 &3399, 3376 &3711, 3562 &5154, 5033    \\
$m_{\tilde{t}_{1,2}}$
                 &1302, 1554  &1181, 1381 &1764, 2115 &2047, 2457 &2469, 2880 &3300, 3957  \\
\hline $m_{ \tilde{d}_{L,R}}$
                 &2250, 2253  &2003, 1996 &2976, 2967 &3398, 3378 &3712, 3561 &5154, 5034   \\
$m_{\tilde{b}_{1,2}}$
                 &1539, 1730  &1366, 1495 &2099, 2333 &2440, 2686 &2791, 2861 &3935, 4137 \\
\hline
$m_{\tilde{\nu}_{1}}$
                 &    2239        &  1955       & 2910     &  3281    & 3059  & 4648 \\
$m_{\tilde{\nu}_{3}}$
                 &    1973        &  1717       & 2575     &  2906    & 2704  & 4135\\
\hline
$m_{ \tilde{e}_{L,R}}$
                &2240, 2235     &1956, 1942     &2911, 2891  &3281, 3246 &3060, 2861 &4648, 4496  \\
$m_{\tilde{\tau}_{1,2}}$
                &1662, 1974     &1424, 1718     &2163, 2574  &2432, 2906 &2017, 2701 &3341, 4131 \\
\hline
$\sigma_{SI}({\rm pb})$
                & $1.3 \times 10^{-8}$ & $1.0 \times 10^{-8}$ & $2.6 \times 10^{-8}$ & $2.6 \times 10^{-8}$ & $1.5 \times 10^{-9}$ &  $1.4 \times 10^{-8}$ \\

$\sigma_{SD}({\rm pb})$
                & $3.7\times 10^{-4}$ & $3.0\times 10^{-5}$ & $7.8\times 10^{-4}$ & $7.8\times 10^{-4}$ & $3.8\times 10^{-7}$ & $5.9\times 10^{-5}$  \\

$\Omega_{CDM}h^2$
               & 0.13  &  0.11  &  0.12 & 0.11 & 0.11 & 0.12 \\
\hline
\hline
\end{tabular}
\caption{Benchmark points for $\tan\beta$=50, $\mu< 0$. All of these points satisfy the various constraints
mentioned in section~\ref{constraintsSection}. For all points, the neutralino
LSP has sizeable higgsino component. The points lie between present and future direct and indirect dark matter search limits
shown in Figures~\ref{SIA0TB50} and \ref{SDA0TB50}. Pt1 shows lightest neutralino and a light gluino. Pt5 shows stau-coannihilation.
}
\label{table6}
\end{table}

\clearpage

\begin{table}
\centering
\begin{tabular}{lccccc}
\hline
\hline
                 & Pt1 & Pt2    & Pt3        \\
\hline
$m_{0}$          & 1771  & 1751   & 4892     \\
$M_{1/2} $       & 686   & 696    & 2075    \\
$A_0 $           & -5000 & -5000  &   0      \\
$sign(\mu) $     &  -   & -     &  -    \\
$\tan\beta$      &  10   &  10    &   53    \\

\hline
$m_h$            &123  & 122    & 122      \\
$m_H$            &2657  & 2650  & 947      \\
$m_A$            &2640  & 2633  & 940    \\
$m_{H^{\pm}}$    &2658  & 2652  & 951     \\

\hline
$m_{\tilde{\chi}^0_{1,2}}$
                 &154, 879  &156, 892   &488, 879    \\
$m_{\tilde{\chi}^0_{3,4}}$
                 &1929, 1932  &1934, 1938  &880, 2696   \\

$m_{\tilde{\chi}^{\pm}_{1,2}}$
                 &886, 1938  &899, 1944  &900, 2658    \\
$m_{\tilde{g}}$  &  1648      & 1669      & 4578     \\

\hline $m_{ \tilde{u}_{L,R}}$
                 &2303, 2217  &2303, 2214 &6398, 6119       \\
$m_{\tilde{t}_{1,2}}$
                 &188, 1695   &196, 1697 &4289 , 5030   \\
\hline $m_{ \tilde{d}_{L,R}}$
                 &2305, 2218 &2304, 2215 &6398,  6119    \\
$m_{\tilde{b}_{1,2}}$
                 &1700, 2144  & 1703, 2141 &4876, 5001   \\
\hline
$m_{\tilde{\nu}_{1}}$
                 &    1889        &  1873       &  5270   \\
$m_{\tilde{\nu}_{3}}$
                 &    1865       &   1849      &  4597     \\
\hline
$m_{ \tilde{e}_{L,R}}$
                &1887, 1775      &1872, 1755    & 5270, 4901  \\
$m_{\tilde{\tau}_{1,2}}$
                &1717, 1864       &1697, 1849   & 3270, 4592    \\
\hline
$\sigma_{SI}({\rm pb})$
                & $1.2 \times 10^{-11}$ & $1.2 \times 10^{-11}$ & $7.8 \times 10^{-10}$  \\

$\sigma_{SD}({\rm pb})$
                & $1.4\times 10^{-9}$ & $2.3\times 10^{-9}$ & $5.2\times 10^{-7}$  \\

$\Omega_{CDM}h^2$
               & 0.12  &  0.13  &  0.1  \\
\hline
\hline
\end{tabular}
\caption{Pts.1 and 2 correspond to stop-neutralino coannihilation region, 
 while Pt3 belongs in the so-called H-funnel region.
}
\label{table7}
\end{table}

\begin{table}
\centering
\begin{tabular}{lcccccccc}
\hline
\hline
                 & NUGM & CMSSM    & NUGM     &  CMSSM & NUGM  & CMSSM & NUGM & CMSSM \\
\hline
$m_{0}$         & 287    & 199     & 538     & 268    & 219    & 150   & 440       &  276          \\
$M_{1/2} $      & 1687    &  918    & 1188    & 650    & 1258   & 691   & 1040      &  572        \\
$A_0 $          & 1000   &  1000   & 1000    & 1000   & 1000   & 1000  & 1000      &  1000         \\
$sign(\mu) $    & -     & -      &  -       & -     &  +    & +      & +      &  +    \\
$\tan\beta$     &  10    &  10     &  30     &  30    &  10    &  10     &  30     &   30    \\

\hline
$m_h$            & 118   & 116     & 118     & 115  & 117  & 115  & 117  & 115     \\
$m_H$            & 2071  & 1160    & 1289    & 726  & 1587 & 888  & 1233 & 687    \\
$m_A$            & 2057  & 1152    & 1281    & 721  & 1576 & 882  & 1225 & 683     \\
$m_{H^{\pm}}$    & 2073  & 1163    & 1292    & 731  & 1589 & 892  & 1236 & 692    \\

\hline
$m_{\tilde{\chi}^0_{1,2}}$
                 &388, 1277     &388, 732   &271, 933   &271, 509   &287, 1005 &287, 538  &236, 840 &236, 442     \\
$m_{\tilde{\chi}^0_{3,4}}$
                 &1281, 2146    &976, 984    &940, 1512  &704, 716   &1011, 1597 &750, 767 &847, 1322 &628, 644  \\

$m_{\tilde{\chi}^{\pm}_{1,2}}$
                 &1298, 2121    &733, 983    &948, 1496  &509, 715   &971, 1574  &539, 767  &811, 1302 &443, 645   \\
$m_{\tilde{g}}$  &  3594        &   2020     &  2601     &1470       & 2736   & 1553 & 2298  & 1307  \\

\hline $m_{ \tilde{u}_{L,R}}$
                  &3474, 3084       &1848, 1776   &2557, 2293  &1364, 1315  &2645, 2357 &1423, 1369  &2252, 2021 &1220, 1178       \\
$m_{\tilde{t}_{1,2}}$
                  &2547, 3285       &1490, 1751   &1875, 2363  &1095, 1273  &2502, 1235 &1145, 1358   &1655, 2102 &975, 1148   \\
\hline $m_{ \tilde{d}_{L,R}}$
                 &3475, 3080       &1850, 1769   &2558, 2290   &1366, 1310   &2646, 2354  &1425, 1364 &2253, 2019 &1223, 1174    \\
$m_{\tilde{b}_{1,2}}$
                 &3065, 3268       &1727, 1760    &2180, 2348  &1230, 1270   &2344, 2487 &1330, 1359  &1960, 2086 &1108, 1151  \\
\hline
$m_{\tilde{\nu}_{1}}$
                 &   1644         &  636       &  1269      &  505    & 1234  & 479  & 1101   & 466   \\
$m_{\tilde{\nu}_{3}}$
                 &   1627         &  628       &  1211      &  478     &1221  & 472  & 1056   & 433    \\
\hline
$m_{ \tilde{e}_{L,R}}$
                &1651, 398        &645, 393      &1275, 576    &513, 362    &1242, 301 &489, 298   &1107, 476   &475, 350   \\
$m_{\tilde{\tau}_{1,2}}$
                &388, 1640        &388, 640      &276, 1220    &276, 496    &290, 286  &290, 484   &241, 1064  &241, 450 \\
$\sigma_{SI}({\rm pb})$
                & $5.4 \times 10^{-11}$ & $5.0 \times 10^{-12}$ & $9.9 \times 10^{-11}$ & $3.7 \times 10^{-11}$ & $7.5 \times 10^{-13}$ & $3.6 \times 10^{-10}$& $3.0 \times 10^{-10}$ & $1.1 \times 10^{-9}$   \\

$\sigma_{SD}({\rm pb})$
                & $3.4\times 10^{-8}$ & $1.0\times 10^{-7}$ & $1.2\times 10^{-7}$  & $4.2\times 10^{-7}$  & $8.5\times 10^{-8}$ & $3.1\times 10^{-7}$ & $7.1\times 10^{-7}$ & $7.0\times 10^{-7}$ \\

$\Omega_{CDM}h^2$
               & 0.13  &  0.13  &  0.12 & 0.08 &0.12 &0.12 &0.11 &0.08    \\
\hline
\hline
\end{tabular}
\caption{
Particle masses for NUGM and CMSSM benchmark points 
 in the stau-neutralino coannihilation region. 
The parameters $m_0$ and $M_{1/2}$ in the CMSSM 
 are tuned so as to yield the same neutralino LSP 
 and ${\tilde{\tau}}$ masses as in the NUGM for each benchmark point. 
We observe that masses of remaining particles can differ widely. 
Results for $\Omega_{CDM}h^2$ in CMSSM are within 5-$\sigma$ 
 range of the WMAP constraints. 
}
\label{table8}
\end{table}

\begin{table}
\centering
\begin{tabular}{lcccccccc}
\hline
\hline
                 & NUGM & CMSSM    & NUGM     &  CMSSM & NUGM  & CMSSM & NUGM & CMSSM \\
\hline
$m_{0}$          & 541   & 423    &   1690  & 1690   & 1700  & 1699   & 1684  & 1683  \\
$M_{1/2} $      & 1203  & 1218    &  201    & 203    & 187   & 188    & 217   & 219     \\
$A_0 $          & 1000  & 1000   &  -1000   & -1000  & 0     & 0       & 0     & 0       \\
$sign(\mu) $    & -    & -     &  +      &  +    & -    & -      & +    & +      \\
$\tan\beta$      & 30   &  30    &   30     & 30     & 30    & 30      &  50   & 50      \\

\hline
$m_h$            & 118  & 119     &  115  & 115      & 114.4 &  115  & 114.4    & 114.5  \\
$m_H$            & 1303 & 1318     & 1472 & 1461     & 1439  &  1444 & 875      & 781    \\
$m_A$            & 1294 & 1310     & 1463 & 1451     & 1429  &  1434 & 869      & 775     \\
$m_{H^{\pm}}$    & 1305 & 1321     & 1475 & 1463     & 1441  &  1446 & 880      & 786    \\

\hline
$m_{\tilde{\chi}^0_{1,2}}$
                 & 275,940    & 523,980   & 44 ,248 &85 ,165 &39 ,172 &77 ,141 & 47,208 & 89,162   \\
$m_{\tilde{\chi}^0_{3,4}}$
                 & 947,1531    & 1243,1252 &408 ,421 &472 ,480 &219 ,295 &254 ,271 & 247,325 & 289,312  \\

$m_{\tilde{\chi}^{\pm}_{1,2}}$
                 & 956,1514    & 981,1252  &247 ,423  &166  ,482 &172 ,294 &139 ,272 & 202,320 & 163,311   \\
$m_{\tilde{g}}$  &  2631       &   2631     &  583    & 583  & 546   & 546  & 617 & 617  \\

\hline $m_{ \tilde{u}_{L,R}}$
                 & 2582,2318    & 2423,2328  &1731 ,1728 &1725 ,1729 &1732 ,1730 &1726 ,1730 & 1736,1730 & 1727,1730      \\
$m_{\tilde{t}_{1,2}}$
                 & 1896,2390    & 1939,2242  &977 ,1349    &964 ,1331 &1010 ,1357 &1012 ,1353 & 1023,1256 & 1025,1232  \\
\hline $m_{ \tilde{d}_{L,R}}$
                  & 2587,2315    & 2423,2315  &1733 , 1729   &1726 ,1729 &1734 ,1731 &1727 ,1731 & 1737,1731 & 1730,1731    \\
$m_{\tilde{b}_{1,2}}$
                  & 2205,2376      & 2202,2236 &1337 ,1594     &1318 ,1583  &1346 ,1540 &1341 ,1597 & 1241,1399 & 1215,1373   \\
\hline
$m_{\tilde{\nu}_{1}}$
                 &  1283          &   905      &  1697      & 1691     & 1705  & 1699  & 1693   & 1685   \\
$m_{\tilde{\nu}_{3}}$
                 &  1226          &   876      &  1625     &  1614     & 1636  & 1631  & 1502   &  1480   \\
\hline
$m_{ \tilde{e}_{L,R}}$
                & 1289,580        & 912,616    &1698 ,1690  &1692 ,1690  &1707 ,1699 &1700 ,1699 & 1695,1683 & 1686,1683   \\
$m_{\tilde{\tau}_{1,2}}$
                & 278,1233       & 541,890   & 1542 ,1627   &1532 ,1616  &1558 ,1638 &1562 ,1633 & 1269,1504 & 1240,1482  \\

\hline
\hline
\end{tabular}
\caption{
Comparison of particle masses of NUGM and CMSSM benchmark points 
 with $m_0$ and $M_{1/2}$ in CMSSM tuned to give 
 the same $m_{\tilde{g}}$ and $m_{ \tilde{d}_{R}}$ as in NUGM.
}
\label{table9}
\end{table}
\end{document}